%% file: ANA-SUSY-2018-04-PAPER.tex
\newcommand*{\ATLASLATEXPATH}{}
\documentclass[PAPER,UKenglish,texlive=2016,cernpreprint]{\ATLASLATEXPATH atlasdoc}

\usepackage[subcaption]{\ATLASLATEXPATH atlaspackage}
\usepackage[BSM]{\ATLASLATEXPATH atlasphysics}
 
\usepackage{\ATLASLATEXPATH atlasbiblatex}
 
\graphicspath{{logos/}{figures/}}
 
\usepackage{ANA-SUSY-2018-04-PAPER-defs}

\AtlasTitle{
Search for direct stau production in events with two hadronic $\tau$-leptons in $\sqrt{s} = 13$ TeV $pp$ collisions with the ATLAS detector
}
 
\author{The ATLAS Collaboration}
 
\AtlasAbstract{
 
A search for the direct production of the supersymmetric partners of $\tau$-leptons (staus)
in final states with two hadronically decaying $\tau$-leptons is presented.
The analysis uses a dataset of $pp$ collisions corresponding to an integrated luminosity of
$139$ fb$^{-1}$, recorded with the ATLAS detector at the Large Hadron Collider
at a center-of-mass energy of 13~TeV.
No significant deviation from the expected Standard Model background is observed.
Limits are derived in scenarios of
direct production of stau pairs with each stau decaying into the stable lightest neutralino and one $\tau$-lepton in simplified models where the two stau mass eigenstates are degenerate. Stau masses from 120 GeV to 390 GeV are excluded at
95\% confidence level for a massless lightest neutralino.
 
}
 
\AtlasRefCode{SUSY-2018-04}
\AtlasJournal{Phys. Rev. D}
\PreprintIdNumber{CERN-EP-2019-191}

\AtlasJournalRef{Phys. Rev. D 101, 032009 (2020)}
\AtlasDOI{10.1103/PhysRevD.101.032009}
 
\hypersetup{pdftitle={ATLAS document},pdfauthor={The ATLAS Collaboration}}
 
\begin{document}
\maketitle
 
\section{Introduction}

Supersymmetry (SUSY)~\cite{Golfand:1971iw,Volkov:1973ix,Wess:1974tw,Wess:1974jb,Ferrara:1974pu,Salam:1974ig,Martin:1997ns}
postulates the existence of superpartners, referred to as \textit{sparticles},
whose spin differs by one half unit from each corresponding Standard Model (SM) particle.
In models that conserve $R$-parity~\cite{Farrar:1978xj},
sparticles are always produced in pairs, and the lightest supersymmetric particle (LSP) is stable and is a dark-matter
candidate~\cite{Goldberg:1983nd,Ellis:1983ew}.
 
In SUSY models, the sector of sparticles with only electroweak interactions contains charginos ($\susy{\chi}^{\pm}_{i}$, $i$ = 1, 2 in order of increasing masses), neutralinos
($\susy{\chi}^{0}_{j}$, $j$ = 1, 2, 3, 4 in order of increasing masses), charged sleptons (\slepton), and sneutrinos ($\susy{\nu}$).
Charginos and neutralinos are the mass eigenstates formed from linear superpositions of the superpartners of the
charged and neutral Higgs bosons and electroweak gauge bosons.
The sleptons are the superpartners of the charged leptons and  are referred to as left or right (\sleptonL or \sleptonR)
depending on the chirality of their SM partners.
The slepton mass eigenstates are a mixture of \sleptonL and \sleptonR, and are labeled as $\susy{\ell}_{k}$ ($k$ = 1, 2 in order of increasing mass).
In this work, the scalar superpartner of the left-handed $\tau$-lepton (the stau-left, $\stau_\mathrm{L}$) and right-handed $\tau$-lepton (the stau-right, $\stau_\mathrm{R}$) are assumed to be mass degenerate.
 
Final states with $\tau$-leptons originating from stau decays are of particular interest for SUSY searches.
Models with light staus can lead to a dark-matter relic density consistent with cosmological observations \cite{Vasquez:2011}
and light sleptons in general could play a role in the co-annihilation of neutralinos~\cite{Belanger:2004ag,King:2007vh}.
Sleptons are expected to have masses of order 100 GeV in gauge-mediated~\cite{Dine:1981gu,AlvarezGaume:1981wy,Nappi:1982hm}
and anomaly-mediated~\cite{Randall:1998uk,Giudice:1998xp} SUSY breaking models.

\begin{figure}[!htb]
\centering
\includegraphics[width=.45\textwidth]{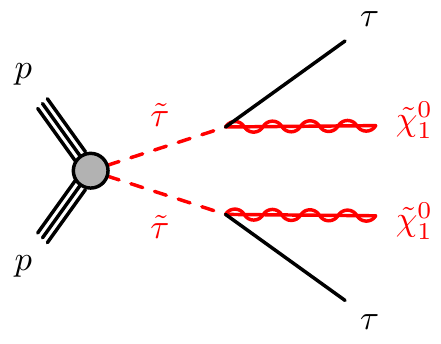}
\caption{A diagram illustrating the pair production of staus and subsequent decay into a two-$\tau$-lepton final state with missing transverse momentum from the neutralinos.}
\label{fig:feynmanslepton_summary}
\end{figure}

In some scenarios the direct production rate of sleptons can greatly exceed the production rate of squarks and gluinos at the Large Hadron Collider (LHC).
In the simplified models studied in this paper, the lightest neutralino is the LSP and is purely the superpartner of the U(1) gauge field (the bino) and not admixed with the superpartner of the SU(2) gauge field (wino).
Further,  the two staus are assumed to be mass-degenerate.  The stau-left and stau-right decay each with a 100\% branching fraction into a bino-like neutralino and a $\tau$-lepton.
All sparticles other than those explicitly mentioned here are assumed to be inaccessible at the LHC energy.
This paper focuses on the direct production of a stau pair, leading to the final state illustrated in Figure~\ref{fig:feynmanslepton_summary}.
 
Signal events are characterized by the presence of exactly two $\tau$-leptons and large missing transverse momentum due to the
undetected neutrinos, as a result of the $\tau$-lepton decays, and lightest neutralinos. Final states with exactly two hadronically decaying $\tau$-leptons ($\tau$ $\rightarrow$ hadrons $\nu_\tau$) are
considered.
 
The search described in this document uses a dataset of $\sqrt{s}=13$~TeV $pp$ collisions collected with the ATLAS detector from 2015 to 2018 at the LHC. 
At LEP~\cite{LEPSUSYWG:01-03.1,alephsleptons, Heister:2003zk, Abdallah:2003xe, L3sleptons, opalsleptons}, searches set a lower limit of 86.6 GeV at 95\% confidence level (CL) on the mass of promptly decaying staus.
Similar searches were performed previously by the ATLAS and CMS Collaborations using data collected at $\sqrt{s} = 8$ TeV in 2012 at the LHC~\cite{SUSY-2014-05, CMS-SUS-14-022}, with the ATLAS result excluding a single $\stau$ mass hypothesis of 109 GeV decaying to a $\tau$-lepton and massless neutralino.
CMS has performed a similar search using an integrated luminosity of 77.2 fb$^{-1}$ of $\sqrt{s}=13$~TeV $pp$ collisions and has excluded degenerate production of $\stau$ pairs with mass up to 150 GeV for a nearly massless neutralino~\cite{CMS-SUS-18-006}.
 
\section{ATLAS detector}
\label{sec:atlas}
 
The ATLAS detector~\cite{PERF-2007-01} is a multipurpose particle physics detector with
forward--backward symmetric cylindrical geometry and nearly $4\pi$
coverage in solid angle.\footnote{
ATLAS uses a right-handed coordinate system with its origin at the nominal
interaction point (IP) in the center of the detector, and the $z$-axis along the
beam line. The $x$-axis points from the IP to the center of the LHC ring,
and the $y$-axis points upwards. Cylindrical coordinates $(r,\phi)$ are used
in the transverse plane, $\phi$ being the azimuthal angle around the $z$-axis.
Observables labeled \textit{transverse} refer to the projection into the $x$--$y$ plane.
The pseudorapidity is defined in terms of the polar angle $\theta$ by $\eta=-\ln\tan(\theta/2)$.
}
It features an inner tracking detector (ID) surrounded by a 2 T superconducting solenoid,
electromagnetic and hadronic calorimeters, and a muon spectrometer (MS).  The ID covers the pseudorapidity region  $|\eta| <2.5$ and consists of a
silicon pixel detector, a silicon microstrip detector, and a transition
radiation tracker. One significant upgrade for the \tttev running period is the presence of the
insertable B-layer~\cite{ATLAS-TDR-2010-19,PIX-2018-001}, an additional pixel layer close to the interaction point  which
provides high-resolution hits at a small radius to improve the tracking and vertex reconstruction performance.
The calorimeters are composed of high-granularity liquid-argon (LAr) electromagnetic calorimeters with lead, copper,
or tungsten absorbers (in the pseudorapidity region $|\eta| < 3.2$) and a steel--scintillator hadronic calorimeter (for
$|\eta|<1.7$).
The endcap and forward regions are instrumented with LAr calorimeters
for both the electromagnetic and hadronic measurements up to $|\eta|=4.9$. The MS surrounds the calorimeters and consists
of three large superconducting air-core toroidal magnets, each with eight coils, a system of precision tracking chambers ($|\eta|<2.7$), and detectors
for triggering ($|\eta|<2.4$).
A two-level trigger system is used to select events for recording~\cite{Aaboud:2016leb}.
\section{Data and simulated event samples}
\label{sec:data}
 
After the application of beam, detector, and data quality
requirements,  the analyzed dataset corresponds to an integrated luminosity of \ourintlumi of $pp$ collision data
recorded from 2015 to 2018  at \tttev.
The uncertainty in the combined 2015--2018 integrated luminosity is 1.7 \% \cite{ATLAS-CONF-2019-021}, obtained using the LUCID-2 detector \cite{LUCID2} for the primary luminosity measurements.
The average number of interactions per bunch crossing (<$\mu$>) for this dataset ranges from about 10 up to 60, with a mean value of 34.
 
Monte Carlo (MC) simulated event samples were used to estimate the SUSY signal yields
and to aid in evaluating the SM backgrounds.
Generated SM events were passed through a detailed detector simulation~\cite{SOFT-2010-01} based on {\GEANT}4~\cite{geant4}.  SUSY MC events were processed with a fast detector simulation that parameterizes the response of the electromagnetic and hadronic calorimeters but uses {\GEANT}4 for the other detectors.
All simulated events were overlaid with multiple $pp$ collisions (pileup) simulated with the soft strong-interaction processes of
\PYTHIA 8.186~\cite{Sjostrand:2007gs} using the A3 set of tuned parameters~\cite{ATL-PHYS-PUB-2016-017} and the NNPDF23LO~\cite{Martin:2009iq} parton distribution function (PDF) set.
The simulated events were reconstructed using the same algorithms as the data, and were reweighted so that the distribution of the expected number of
collisions per bunch crossing matched the one in the data.
 
\subsection{Simulated background samples}
\label{sec:SM}
 
Events with $Z/\gamma^*\to\ell\ell$ $(\ell=e,\mu,\tau)$ and $W\to\ell\nu$ produced with accompanying jets (including jets initiated by heavy flavor quarks) were generated with \SHERPA 2.2.1~\cite{Gleisberg:2008ta,Hoeche:2012yf}.
Matrix elements (ME) were calculated for up to two additional partons at NLO and four additional partons at leading order (LO), using the Comix~\cite{Gleisberg:2008fv} and \openloops~\cite{Cascioli:2011va,Denner:2016kdg} generators and merged with the \SHERPA parton shower (PS)~\cite{Schumann:2007mg} using the ME+PS@NLO prescription~\cite{Hoeche:2012yf}.
The NNPDF3.0NNLO~\cite{Ball:2014uwa} PDF set was used in conjunction with a dedicated PS tuning developed by the \SHERPA authors.
The $W/Z$+jets events were normalized using their next-to-next-to-leading-order (NNLO) cross sections~\cite{Catani:2009sm}.

Fully leptonically and semileptonically decaying diboson samples ($VV=WW/WZ/ZZ$)  were simulated with the
\SHERPA~2.2.1 and 2.2.2~\cite{Gleisberg:2008ta} generator at NLO. In this setup, multiple
matrix elements were matched and merged with the \SHERPA parton shower
based on Catani--Seymour dipole factorization~\cite{Schumann:2007mg}
using the MEPS@NLO
prescription~\cite{Hoeche:2011fd,Hoeche:2012yf,Catani:2001cc,Hoeche:2009rj}.
The virtual QCD corrections for matrix elements at NLO accuracy were
provided by the \openloops library~\cite{Denner:2016kdg}. Samples were generated using the
NNPDF3.0NNLO set, along with the dedicated set of
tuned parton-shower parameters developed by the \SHERPA authors.
 
The production of top-quark pairs (\ttbar) and single top quarks in the $Wt$ and $s$-channels was performed with \POWHEG-\textsc{Box} v2~\cite{Frixione:2007nw,Nason:2004rx,Frixione:2007vw,Alioli:2010xd}, with the NNPDF2.3LO~\cite{Ball:2012cx} PDF set at NLO
in the ME calculations and the ATLAS underlying-event tune A14~\cite{ATL-PHYS-PUB-2014-021}.
Electroweak $t$-channel single-top-quark events were generated using the \POWHEG-\textsc{Box} v2 event generator.
The PS, fragmentation, and the underlying event were simulated using \PYTHIA 8.186 with the NNPDF2.3LO PDF set and a corresponding set of A14 tuned parameters.
The top-quark mass was set to 172.5 GeV.
The \ttbar\ sample is normalised to the cross-section prediction at next-to-next-to-leading order (NNLO)
in QCD including the resummation of next-to-next-to-leading logarithmic (NNLL) soft-gluon terms calculated using
\textsc{Top++2.0}~\cite{Beneke:2011mq,Cacciari:2011hy,Baernreuther:2012ws,Czakon:2012zr,Czakon:2012pz,Czakon:2013goa,Czakon:2011xx}.
The cross section for single-top-quark was computed
for the $Wt$-channel at NLO in QCD with NNLL soft gluon
corrections~\cite{Aliev:2010zk,Kant:2014oha}, and to NLO in QCD for the $t$- and $s$-channels~\cite{Aliev:2010zk,Kant:2014oha}.
Top-quark pair production with an additional $W$ or $Z$ boson was calculated using \MGMCatNLO  2.2.2~\cite{Alwall:2014hca} at NLO in the ME calculations, while fragmentation and hadronization
were simulated with \PYTHIA 8.186.
The underlying-event tune A14 was used with the
NNPDF2.3LO PDF set, and the cross sections were normalized using NLO predictions~\cite{Lazopoulos:2008de,Campbell:2012dh}.
 
Small contributions from Higgs boson events produced by gluon--gluon fusion and vector-boson fusion were modeled using \POWHEG-\textsc{Box} v2 with the NNPDF3.0NNLO PDF and showered using \PYTHIA 8.186.  Contributions from the associated production of a Higgs boson with a vector boson and from a Higgs boson in association with two top quarks were simulated using \PYTHIA 8.186 and \MGMCatNLO, respectively.  All Higgs boson samples were normalized to cross sections from Ref.~\cite{deFlorian:2016spz}.
 
For all samples showered with \PYTHIA 8, EvtGen 1.2.0~\cite{EvtGen} was used to simulate the decays of bottom and charmed hadrons.
 
\subsection{Simulated signal samples}
\label{sec:SUSY}
 
Simulated signal samples were generated using \MGMCatNLO 2.6.2 interfaced to \PYTHIA 8.186 with the A14 tune for the PS modeling,
hadronization, and underlying event. The ME calculation was performed at tree level and includes the emission of up to two additional partons.
The NNPDF2.3LO PDF set was used. The ME--PS matching used the CKKW-L~\cite{CKKW} prescription, with a matching scale set to one quarter of the mass of the pair of produced particles.
Signal cross sections were calculated to next-to-leading order in the strong
coupling constant, adding the resummation of soft gluon emission at
next-to-leading-logarithm accuracy (NLO+NLL) using the PDF4LHC15mc PDF set~\cite{Fuks:2013vua}.
The nominal cross section and its uncertainty were taken from an envelope of cross section predictions
using different PDF sets and factorization and renormalization scales, as described in Refs.~\cite{Beenakker:1999xh,Bozzi:2007qr,Fuks:2013vua,Fuks:2013lya,Fiaschi:2018xdm}.
 
The masses of all charginos and neutralinos apart from the \ninoone, were set to 2.5 TeV, thus leaving a single
kinematically allowed decay: $\tilde{\tau} \to $ \ninoone\ $\tau$.
The stau-left and stau-right were combined and have the same mass, which was varied between 100 and 440 GeV and no mixing was assumed between the gauge and mass eigenstates.  The mass of the bino-like \ninoone\ was varied in the range 1--200~GeV.
 
Reference points with $\stau$ masses of 120 GeV, 280 GeV, and a \ninoone  mass of 1 GeV are used throughout this paper to illustrate typical features of the SUSY models to which this analysis is sensitive. The theoretical cross section used at NLO+NLL was 140 (50) fb with $\stau_\mathrm{L} \stau_\mathrm{L}$ ($\stau_\mathrm{R} \stau_\mathrm{R}$) of 120 GeV, and 5.8 (2.2) fb  with $\stau_\mathrm{L} \stau_\mathrm{L}$ ($\stau_\mathrm{R} \stau_\mathrm{R}$) of 280 GeV.
\section{Event reconstruction}
\label{sec:object}
 
\newcommand{\pTvec}{\mathbf{p}_\mathrm{T}}
\newcommand{\qTvec}{\mathbf{q}_\mathrm{T}}
\newcommand{\qT}{q_\mathrm{T}}
\newcommand{\pTell}[1]{\mathbf{p}_\mathrm{T}^{\ell#1}}
\newcommand{\pTmiss}{\mathbf{p}_\mathrm{T}^\mathrm{miss}}
 
Events with at least one reconstructed primary vertex~\cite{ATL-PHYS-PUB-2015-026} are selected. A primary vertex must have
at least two associated charged-particle tracks with transverse momentum $p_\text{T} >$~500~MeV and be consistent with the collision region.
In events with multiple primary vertices, the one with the largest $\sum\pt^2$ of the associated
tracks is chosen.
 
Jets are reconstructed from three-dimensional calorimeter energy clusters~\cite{PERF-2014-07} using the anti-$k_t$ algorithm~\cite{anti-kt,Cacciari:2011ma}
with a radius parameter of 0.4. Jet energies are corrected for detector inhomogeneities,
the non-compensating response of the calorimeter, and the impact of pileup, using factors derived from test beam
and $pp$ collision data, and from a detailed {\GEANT}4 detector simulation \cite{PERF-2016-04}.
The impact of pileup is accounted for using a technique, based on jet areas, that provides an
event-by-event and jet-by-jet correction \cite{jetarea}.
Jets that are likely to have originated from pileup are not considered in this analysis~\cite{ATLAS-CONF-2014-018}.  The efficiency of this pileup-rejection selection is approximately 92\%.
Jets are required to have $p_\text{T} >20$~GeV and $\vert\eta\vert < 2.8$.
Events containing jets that are likely to have arisen from detector noise or cosmic rays are removed.
 
Jets containing $b$-hadrons (\bjets) are identified using the MV2c10 algorithm~\cite{PERF-2016-05}, based on a multivariate discriminant making use of track impact parameters and
reconstructed secondary vertices~\cite{FTAG-2018-01}.  Candidate \bjets are required to have $p_\text{T} > 20$~GeV and $|\eta| < 2.5$.  A working point with an average $b$-tagging efficiency of 77\% for simulated \ttbar events is used.  This working point corresponds to a $c$-jet and light-jet rejection of 4.9 and 110, respectively~\cite{ATLAS-CONF-2018-001,ATLAS-CONF-2018-006}.
 
Electron candidates are reconstructed by matching clusters in the electromagnetic calorimeter
with charged-particle tracks in the inner detector. Electrons are required to have $p_\text{T} >17~\GeV$~and $|\eta|<2.47$, and to satisfy the
`loose' working point according to a likelihood-based identification~\cite{Photons}.
Muon candidates are reconstructed from MS tracks matching
ID tracks.
Muons are required to have $p_\text{T} > 14~\GeV$ and $\vert\eta\vert < 2.7$ and fulfill the `medium' quality criteria of Ref.~\cite{PERF-2015-10}.
Events containing a muon candidate with a poorly measured charge-to-momentum ratio ($\sigma(q/p)\,/\,|q/p|>0.4$) are rejected.
 
Electrons and muons are required to satisfy isolation criteria to reduce the number of jets misidentified as charged leptons. The scalar sum of the  $p_\text{T}$  of tracks inside a variable-size cone around the lepton (excluding its own track), must be less than 15\% of the lepton $p_\text{T}$. The track isolation cone size\footnote{$\Delta R = \sqrt{(\Delta y)^2 + (\Delta\phi)^2}~$, where $y$ is the rapidity and $\phi$ the azimuthal angle.} for electrons (muons) $\Delta R$ is given by the minimum of $\Delta R = 10$ GeV/ $p_\text{T}$ and $\Delta R$ = 0.2 (0.3).
In addition, for electrons (muons) the sum of the transverse energy of the calorimeter energy clusters in a cone of $\Delta R$ = 0.2 around the lepton, excluding the energy from the lepton itself, must be less than 20\% (30\%) of the lepton $p_\text{T}$.
For electrons with $p_\text{T} >$ 200 GeV these isolation requirements are not applied, and instead an upper limit of max(0.015 × $p_\text{T}$ [GeV], 3.5 GeV) is placed on the transverse energy of the calorimeter energy clusters in a cone of  $\Delta R$ = 0.2 around the electron.
 
The efficiencies for electrons and muons to satisfy the reconstruction, identification, and isolation criteria are measured in samples of leptonic $Z$ and $J/\psi$ decays, and corrections are applied to the simulated samples to reproduce the efficiencies in data.
 
The reconstruction of hadronically decaying $\tau$-leptons is based on information from tracking in the ID and three-dimensional clusters of energy in the electromagnetic and hadronic calorimeters.
The $\tau$-lepton reconstruction algorithm is seeded by jets reconstructed as described above but with $p_{\mathrm T}>10~\GeV$ and $|\eta|<2.5$.
The reconstructed energies of the hadronically decaying $\tau$-lepton candidates are corrected to the $\tau$-lepton energy scale, which is calibrated using simulation and in situ measurements using $Z \rightarrow \tau \tau$ decays~\cite{ATLAS-CONF-2017-029}.
The $\tau$-neutrino from the hadronic $\tau$-lepton decay is not taken into account in the reconstruction and calibration of the $\tau$-lepton energy and momentum.
Hadronic $\tau$-lepton decay candidates are
required to have one or three associated charged-particle tracks (prongs) and the total electric charge of those tracks must be $\pm 1$ times the electron charge.
To improve the discrimination between hadronically decaying $\tau$-leptons and jets a multivariate algorithm is used
\cite{ATLAS-CONF-2017-029}. The $\tau$-lepton identification algorithm is based on a boosted decision tree (BDT) method.
The BDT algorithm uses various track and cluster variables as input to discriminate between $\tau$-leptons and jets.
For 1-prong (3-prong) $\tau$-lepton candidates, the signal efficiencies are 75\% (60\%) and 60\% (45\%)
for the `medium' and `tight' working points, respectively.
In the following, $\tau$-lepton candidates are required to satisfy the `medium' identification criteria for jet discrimination (`medium' $\tau$-lepton candidates), unless otherwise stated.
For electron discrimination,  an additional BDT is trained to discriminate $\tau$-leptons from electrons and a selection is applied on the output of the discriminant.
This requirement has about 95\% efficiency for selecting $\tau$-leptons, and a background electron rejection factor from 10 to 50 depending on the $\eta$.
The $\tau$-lepton candidates are required to have $p_\text{T} > 20~\GeV$ and $\vert\eta\vert < 2.5$, excluding the transition region between the barrel and endcap calorimeters ($1.37 < |\eta| < 1.52$).
 
The simulation is corrected for differences between the efficiencies of $\tau$ identification in data and simulation at both trigger and reconstruction level.
For hadronically decaying $\tau$-leptons originating from prompt gauge-boson decays, the corrections are calculated with a \textit{tag-and-probe} method in a sample of
$Z \rightarrow \tau \tau$ events where one $\tau$-lepton decays hadronically and the other leptonically into a muon and two neutrinos~\cite{PERF-2013-06}.
Small differences between data and simulation for electron, muon, jet identification, and trigger efficiencies are corrected with scale factors derived in control regions.  For the trigger corrections, dedicated regions enriched in $t\bar{t}$, $Z$, and $W$ events are used.
 
The measurement of the missing transverse momentum vector, $\pTmiss$, and its magnitude, $\met$, is based on the negative vectorial sum of the
$\mathbf{p_\text{T}}$ of all identified jets, $\tau$-lepton candidates, electrons, muons, photons~\cite{Photons}, and an additional soft term. The soft term is constructed from all high-quality tracks
that are associated with the primary vertex but not with any identified particle or jet. In this way, the missing transverse momentum is adjusted for the best
calibration of the jets and the other identified particles, while maintaining pileup independence in the soft term~\cite{PERF-2016-07}.
 
Possible double counting of reconstructed objects is resolved in the following order.
The $\tau$-lepton candidates close to electron or muon candidates ($\Delta R < 0.2$) are
removed, as are electrons that share a track with a muon.  For electrons close to a
jet ($\Delta R < 0.4$), the electron is removed, except when $\Delta R < 0.2$ and the
jet is not $b$-tagged, in which case the jet is removed. For a muon close to a jet ($\Delta R < 0.4$), the muon is removed unless the jet has less than three tracks associated with it and is within $\Delta R < 0.2$.  In the latter case, the jet is removed.
Any remaining jet within $\Delta R = 0.2$ of a $\tau$-lepton candidate is removed.
\section{Event selection}
\label{sec:SR}
 
The events are required to have exactly two
`medium' $\tau$-lepton candidates with opposite-sign electric charge (OS) and to have passed either an
\textit{asymmetric di-$\tau$} trigger or a combined \textit{di-$\tau$ +
\met} trigger~\cite{ATLAS-CONF-2017-061}.
In events selected by the di-$\tau$ + \met trigger, the offline
reconstructed \met must be larger than 150 GeV, to achieve full trigger efficiency.
In 2015--2017 data the efficiency of the asymmetric di-$\tau$ (di-$\tau$ + \met) trigger is 75 -- 80\%
for events with the leading $\tau$-lepton candidate $p_\text{T}$ $>$
95 (50) GeV and the subleading $\tau$-lepton candidate $p_\text{T}$
$>$ 60 (40) GeV. The efficiency is evaluated for events with correctly
identified $\tau$-leptons and the $p_\text{T}$ thresholds are quoted
at reconstruction level. As the luminosity increased in later years,
the trigger selection was tightened; in 2018 the same efficiency was
reached for the leading $\tau$-lepton candidate $p_\text{T}$ $>$ 95
(75) GeV and the subleading $\tau$-lepton candidate $p_\text{T}$ $>$
75 (40) GeV for the asymmetric di-$\tau$ (di-$\tau$ +
\met) trigger.
Events with an additional third `medium' $\tau$-lepton or light lepton are rejected.
 
The reconstructed invariant mass of the visible decay products of the two leading $\tau$-lepton candidates, $m(\tau_1,\tau_2)$, must be
larger than 120~GeV to remove $\tau$-leptons originating from decays of
low-mass resonances and to suppress contributions from $Z$+jets and Higgs boson events ($Z/H$-veto).
 
In order to further discriminate between SUSY signal events and SM background processes, additional requirements are applied to define
the signal region (SR) selections.
To reject events from SM processes containing a top quark, selected events must not contain any $b$-tagged jets ($b$-jet veto).
A lower bound on the \textit{stransverse mass} \mttwo~\cite{Lester:1999tx,Barr:2003rg} is imposed to reduce contributions from \ttbar\ and $WW$ events.
The \mttwo\ variable is defined as:
\[
\mttwo = \min_{\qTvec}\left[\max\left(m_{{\mathrm T}, \tau_1}(\mathbf{p}_{\mathrm{T}, \tau_1},\qTvec),m_{{\mathrm T}, \tau_2}(\mathbf{p}_{\mathrm{T}, \tau_2},\pTmiss-\qTvec)\right)\right],
\]
where $\mathbf{p}_{\mathrm{T}, \tau_1}$ and $\mathbf{p}_{\mathrm{T}, \tau_2}$ are the transverse momenta of the two $\tau$-lepton candidates,
and the transverse momentum vector of one of the invisible particles,
$\qTvec$, is chosen to minimize the larger of the
two transverse masses $m_{{\mathrm T}, \tau_1}$ and $m_{{\mathrm T}, \tau_2}$.
The latter masses are defined by
\[
m_{{\mathrm T}}(\pTvec,\qTvec) = \sqrt{2(\pT\qT-\pTvec\cdot\qTvec)}.
\]
For \ttbar\ and $WW$ events, in which two $W$ bosons decay leptonically and $\pTmiss$
is the sum of the transverse momenta of the two neutrinos, the \mttwo\ distribution has a kinematic endpoint at the $W$ boson mass.
For large mass differences between the staus and the lightest neutralino, the \mttwo\
distribution for signal events extends significantly beyond this endpoint.

The SRs were optimized for stau discovery by varying the kinematic selection criteria.
Events are selected by the asymmetric di-$\tau$
trigger to cover the low stau mass region (SR-lowMass) and by
the di-$\tau$+\met trigger to cover the high stau mass region (SR-highMass), as shown in Table~\ref{tab:signalregions}.
Values of $\met <$ 150 GeV ($\met >$ 150 GeV) are required for SR-lowMass
(SR-highMass) to keep the two SRs orthogonal.
Both (at least one) of the $\tau$-lepton candidates must satisfy the tight
identification criteria for jet discrimination (`tight' $\tau$-lepton candidate)
for SR-lowMass (SR-highMass). Values of $\met >$ 75 GeV are required for
SR-lowMass to increase signal sensitivity.
In addition, the two $\tau$-lepton candidates are required to satisfy $\Delta R(\tau_1,\tau_2) <$
3.2, $|\Delta \phi(\tau_1,\tau_2)| >$ 0.8 and $\mttwo >$ 70 GeV to further suppress contributions from SM
backgrounds.
The product of acceptance and efficiency is around
0.04\% (0.03\%) and 0.53\% (1.22\%) for the reference points with $\stau$ masses of 120
GeV and 280 GeV, respectively, and a \ninoone  mass of 1 GeV in SR-lowMass (SR-highMass).
 
\begin{table}[!htbp]
\caption{Summary of selection requirements for the signal regions.
\label{tab:signalregions}}
\small
\centering
\begin{tabular}{c|c}
\toprule
SR-lowMass  & SR-highMass   \\
\hline
\hline
2 tight $\tau$ (OS) & 2 medium $\tau$ (OS) , $\ge$ 1 tight $\tau$ \\
asymmetric di-$\tau$ trigger  & di-$\tau$+\met trigger \\
75 $< \met <$ 150 GeV    & $\met >$ 150 GeV  \\
\multicolumn{2}{c}{$\tau$ $p_\text{T}$ cut described in Section \ref{sec:SR}} \\
\multicolumn{2}{c}{light lepton veto and 3rd medium $\tau$ veto} \\
\multicolumn{2}{c}{$b$-jet veto}  \\
\multicolumn{2}{c}{$Z/H$ veto ($m (\tau_{1},\tau_{2}) >$ 120 GeV)} \\
\multicolumn{2}{c}{$|\Delta \phi(\tau_1,\tau_2)|>$ 0.8}  \\
\multicolumn{2}{c}{$\Delta R(\tau_1,\tau_2) <$ 3.2} \\
\multicolumn{2}{c}{$\mttwo>$ 70 GeV}\\
\hline
\bottomrule
\end{tabular}
\end{table}
\section{Standard Model background estimation}\label{sec:backgrounds}
 
The main SM processes contributing to the selected final states are
multi-jet, $W$+jets and diboson production.
Background events may contain a combination of `real' $\tau$-leptons, defined as
correctly identified prompt $\tau$-leptons, or `fake' $\tau$-leptons, which can originate from a
misidentified quark or gluon jet, an electron, or a muon.
 
In multi-jet events in the SRs, nearly all $\tau$-lepton candidates are misidentified jets.
The multi-jet contribution in the SRs is estimated from data, as described in Section
\ref{sec:BG_QCD}.
The contribution arising from heavy-flavor multi-jet events containing
a real $\tau$-lepton from the heavy-flavor quark decay is included in the
multi-jet estimation.
The contribution of $W$+jets events, which contain one real $\tau$-lepton from
the $W$ boson decay and one or more misidentified jets, is estimated from MC
simulation and normalized to data in a dedicated control region (CR), as described in Section \ref{sec:BG_W}.
 
Diboson production contributes mainly events containing real $\tau$-leptons originating from
$WW$ and $ZZ$ decaying into a $\tau\tau\nu\nu$ final state.
Additional SM backgrounds arise from $Z$+jets production, or events
that contain a top quark or a top-quark pair in association with jets or
additional $W$ or $Z$ bosons (collectively referred to as \textit{top}-quark background in the following).
The contribution from real $\tau$-leptons exceeds 90\% in
$Z$+jets and diboson production, and ranges from 45\% to 75\% in backgrounds containing
top quarks according to the MC simulation. The contribution of fake $\tau$-leptons
from heavy-flavor decays in jets is found to be negligible in MC
simulation.
To estimate the irreducible background, which includes diboson,
$Z$+jets and top-quark events, only MC simulated samples are used
and validated in dedicated validation regions (VRs), as
described in Section \ref{sec:BG_irre}.

A simultaneous fit of the event yields in the control regions, based on the profile likelihood method, is performed to normalize
the multi-jet and $W$+jets background estimates and propagate systematic uncertainties, as described in Section \ref{sec:fit}.
The sources of systematic uncertainty in the background estimates are described in Section \ref{sec:Systematics}.
Results showing the level of agreement in the validation regions after the fit
are labeled ``post-fit'' and results shown in the control regions prior to the fit are labeled ``pre-fit.''

\subsection{Multi-jet background estimation}
\label{sec:BG_QCD}
 
\begin{figure}[!htb]
\centering
\includegraphics[width=0.48\textwidth]{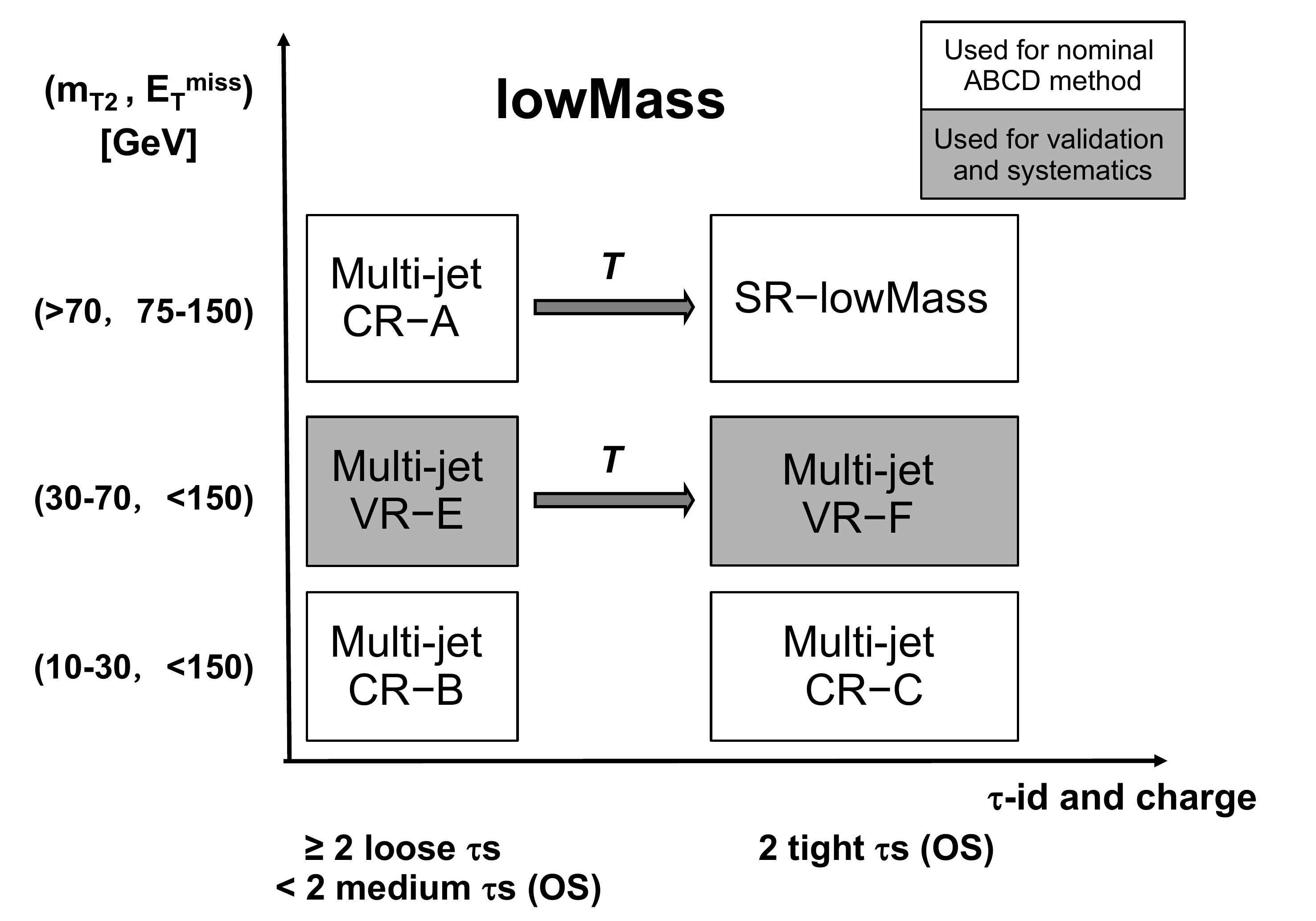}
\includegraphics[width=0.48\textwidth]{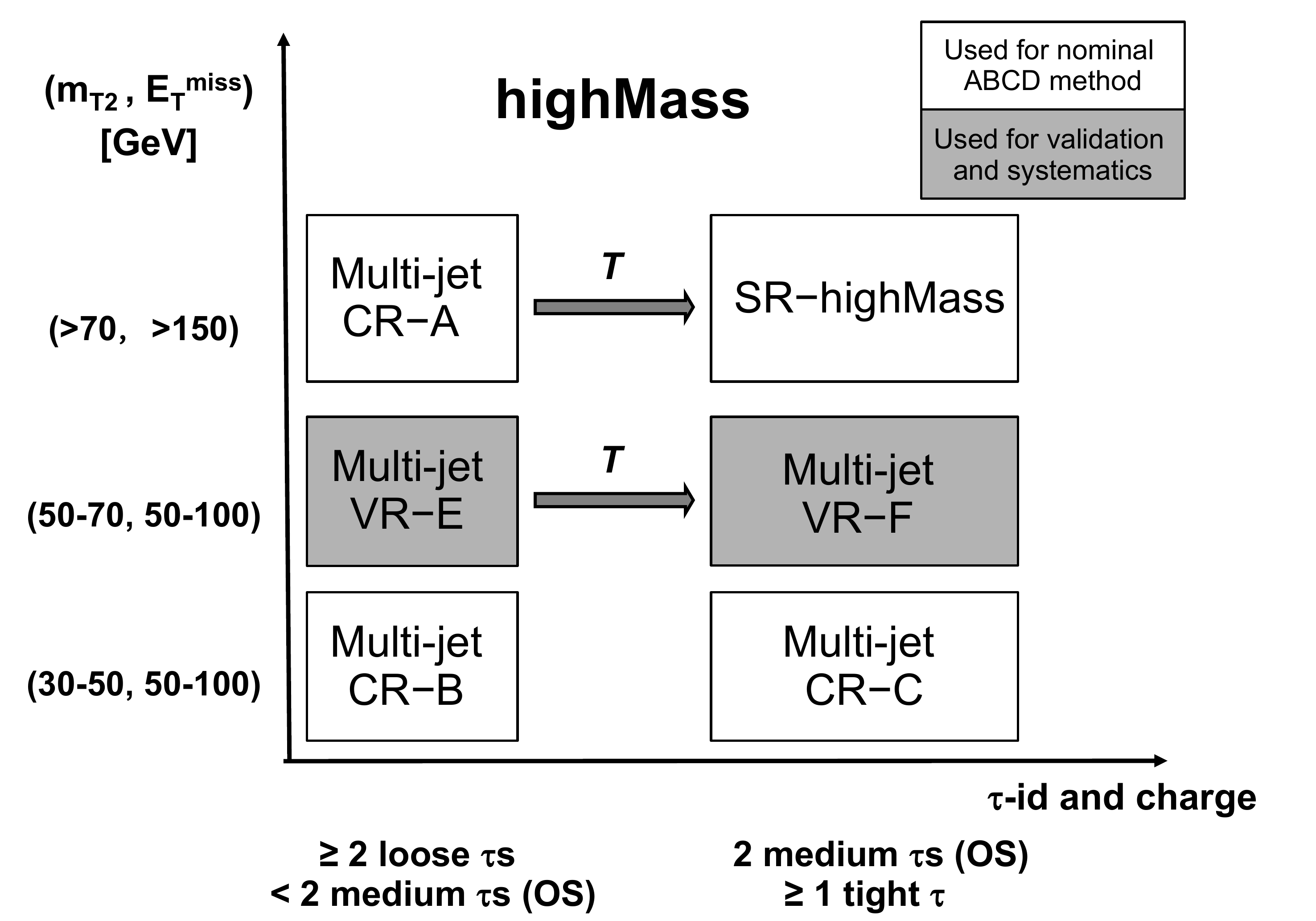}
\caption{ Illustration of the ABCD method for the multi-jet background
determination for SR-lowMass (left) and SR-highMass (right).
The control regions A, B, C, and signal region D for the ABCD method described in the text
(labeled as CR-A, CR-B, CR-C and SR-lowMass/SR-highMass) are drawn as white boxes.
Shown shaded and labeled VR are the regions E and F,
which are used to validate the ABCD method and to estimate the
systematic uncertainty. The transfer factor
\textit{$T$} used in the ABCD method is the ratio of number of multi-jet events
in the regions C and B.
}
\label{fig:abcd}
\end{figure}

One of the dominant backgrounds in the SRs originates from jets
misidentified as $\tau$-leptons in multi-jet production. It accounts for
44\% (30\%) of the
total SM contribution in SR-lowMass (SR-highMass).
This contribution is estimated from data using the so-called
\textit{ABCD method}.
All regions used for the ABCD method are schematically drawn in Figure \ref{fig:abcd}.
Four exclusive regions, labeled as A, B, C, and D, are defined in a two-dimensional plane
as a function of two (or more) discriminating variables that are
largely uncorrelated.
The ratio of events in the regions C and B is then equal to that in the regions D and A.
The number of multi-jet events in region D, $N_{\mathrm D}$, can
therefore be calculated from the multi-jet events in region A, $N_{\mathrm A}$,
multiplied by the transfer factor
\textit{$T=N_{\mathrm C}/N_{\mathrm B}$},
where
$N_{\mathrm C}$ ($N_{\mathrm B}$) is the number of multi-jet events
in region C (B).
The region D corresponds to the SR defined in Section \ref{sec:SR},
whereas the regions A, B, and C are control regions defined accordingly.
In the following, the regions A, B, C, D are labeled as CR-A, CR-B,
CR-C and SR-lowMass (or SR-highMass), respectively. The ABCD method only provides a
first-order estimate of multi-jet background, the normalization and
uncertainty being then modified by a combined fit to CR-A (lowMass)
and CR-A (highMass) described in Section \ref{sec:fit}, in which the transfer
factor is used as an input and fixed. The above two-fold method has
the advantages of taking the possibility of contamination from signal in multi-jet CR-A into
account in the signal exclusion fit described in
Section \ref{section:interpretation}, as well as considering the
correlation of systematic uncertainties among control regions and among background processes.

The definition of the regions used in the ABCD method for
the multi-jet estimation is given in Table~\ref{tab:QCD-CR}; only
those requirements that differ between the CRs/VRs and the SRs are listed.
In all of the regions, the di-$\tau$ +
\met or asymmetric di-$\tau$ triggers described in Section
\ref{sec:SR} are used.
CR-A and CR-B include the events with two `loose' $\tau$-leptons, either
same sign (SS) or OS, and veto the events with two `medium' OS $\tau$-leptons to remain
orthogonal to the SR and reduce the potential contamination by signal events. CR-C and CR-B have events with lower \mttwo~and
\met~; no $\Delta R(\tau_1,\tau_2)$ requirement is applied.
Furthermore, two sets of VRs, VR-E and VR-F,  are defined corresponding to each SR.
The validation regions are used to validate the extrapolation of the ABCD estimation to the SRs and to estimate the systematic
uncertainty from the residual correlation between the $\tau$-lepton identification and charge
requirements and the kinematic variable \mttwo.
 
\def \hfillx {\hspace*{-\textwidth} \hfill}
 
\begin{table}[ht!]
\caption{Definition of the regions used in the ABCD method for the
multi-jet estimation in the SRs. Only
those requirements that differ between the CRs/VRs and the SRs are listed. }
\smallskip
\centering
\footnotesize
 
\begin{minipage}{0.45\textwidth}
\smallskip
 
\begin{tabular}{c|c}
\toprule
$\mathbf {CR-A~(lowMass)}$    & $\mathbf {SR-lowMass~}$\\
\hline
$\ge$ 2 loose $\tau$       & 2 tight $\tau$ (OS)   \\
$<$ 2 medium $\tau$ (OS)    &      --      \\
$\Delta R(\tau_1,\tau_2) <$ 3.2   &     $\Delta R(\tau_1,\tau_2) <$ 3.2    \\
$75$ $<$ $\met$ $< 150$ GeV         &      $75$ $<$ $\met$ $< 150$ GeV    \\ 
$\mttwo >$ 70 GeV             & $\mttwo >$ 70 GeV           \\ 
\hline
\hline
$\mathbf {VR-E~(lowMass)}$    & $\mathbf {VR-F~(lowMass)}$\\
\hline                                                      \hline
$\ge$ 2 loose $\tau$       & 2 tight $\tau$ (OS)   \\
$<$ 2 medium $\tau$ (OS)    &      --      \\
$\Delta R(\tau_1,\tau_2) <$ 3.2   &     $\Delta R(\tau_1,\tau_2) <$ 3.2    \\
$\met$ $< 150$ GeV         &     $\met$ $< 150$ GeV    \\ 
30 $<\mttwo <$~70 GeV         & 30 $<\mttwo <$~70 GeV       \\ 
\hline
\hline
$\mathbf {CR-B~(lowMass)}$    & $\mathbf {CR-C~(lowMass)}$ \\
\hline                                                      \hline
$\ge$ 2 loose $\tau$       &  2 tight $\tau$ (OS)   \\
$<$ 2 medium $\tau$ (OS)    &      --      \\
no $\Delta R(\tau_1,\tau_2) $ cut   &           no $\Delta R(\tau_1,\tau_2) $ cut                   \\
$\met$ $< 150$ GeV         &     $\met$ $< 150$ GeV    \\ 
10 $<\mttwo <$~30 GeV         & 10 $<\mttwo <$~30 GeV       \\ 
\hline
\bottomrule
 
\end{tabular}
 
\end{minipage}
\begin{minipage}{0.45\textwidth}
\smallskip
 
\begin{tabular}{c|c}
\toprule
$\mathbf {CR-A~(highMass)}$     & $\mathbf {SR-highMass~}$\\
\hline
$\ge$ 2 loose $\tau$       & 2 medium $\tau$ (OS)   \\
$<$ 2 medium $\tau$ (OS)   &   $\ge$ 1 tight $\tau$        \\
$\met$ $> 150$ GeV         &      $\met$ $> 150$ GeV    \\ 
$\Delta R(\tau_1,\tau_2) <$ 3.2   &        $\Delta R(\tau_1,\tau_2) <$ 3.2                     \\
$\mttwo >$ 70 GeV             & $\mttwo >$ 70 GeV           \\ 
\hline
\hline
$\mathbf {VR-E~(highMass)}$   & $\mathbf {VR-F~(highMass)}$ \\
\hline                                                      \hline
$\ge$ 2 loose $\tau$       &  2 medium $\tau$ (OS)    \\
$<$ 2 medium $\tau$ (OS)   &    $\ge$ 1 tight $\tau$          \\
$50 < \met < 100$ GeV         & $50 < \met < 100$ GeV  \\ 
$\Delta R(\tau_1,\tau_2) <$ 3.2   &        $\Delta R(\tau_1,\tau_2) <$ 3.2                     \\
50 $<\mttwo <$~70 GeV         & 50 $<\mttwo <$~70 GeV       \\ 
\hline
\hline
$\mathbf {CR-B~(highMass)}$     & $\mathbf {CR-C~(highMass)}$ \\
\hline                                                      \hline
$\ge$ 2 loose $\tau$       &   2 medium $\tau$ (OS)   \\
$<$ 2 medium $\tau$ (OS)     &     $\ge$ 1 tight $\tau$         \\
no $\Delta R(\tau_1,\tau_2) $ cut   &           no $\Delta R(\tau_1,\tau_2) $ cut                   \\
$50 < \met < 100$ GeV         & $50 < \met < 100$ GeV  \\ 
30 $<\mttwo <$~50 GeV         & 30 $<\mttwo <$~50 GeV       \\ 
\hline
\bottomrule
 
\end{tabular}
\end{minipage}
\label{tab:QCD-CR}
\end{table}

The number of multi-jet events in each control region and validation region is estimated from data after
subtraction of other SM contributions estimated from MC simulation.
In CR-B and VR-E, respectively, around 96\% and 90\% (75\% and 79\%) of the events
are from multi-jet production in the lowMass (highMass) region.
For CR-A and CR-C, respectively, the multi-jet purity is 74\% and 57\% (58\% and 53\%) in the lowMass (highMass) regions.
In VR-F the multi-jet purity is 51\%.
 
\begin{figure}[htpb]
\centering
\includegraphics[width=0.495\textwidth]{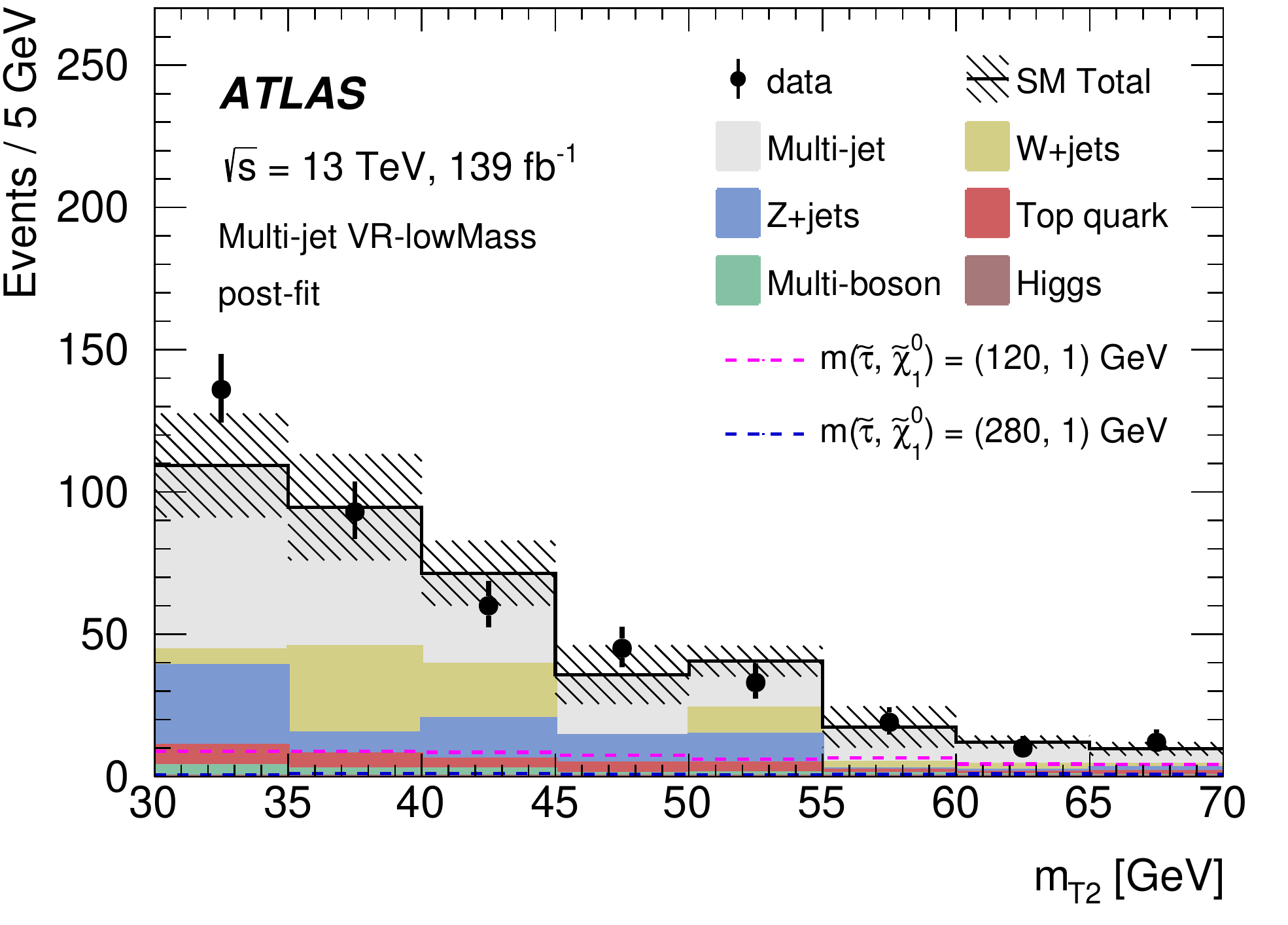}
\includegraphics[width=0.495\textwidth]{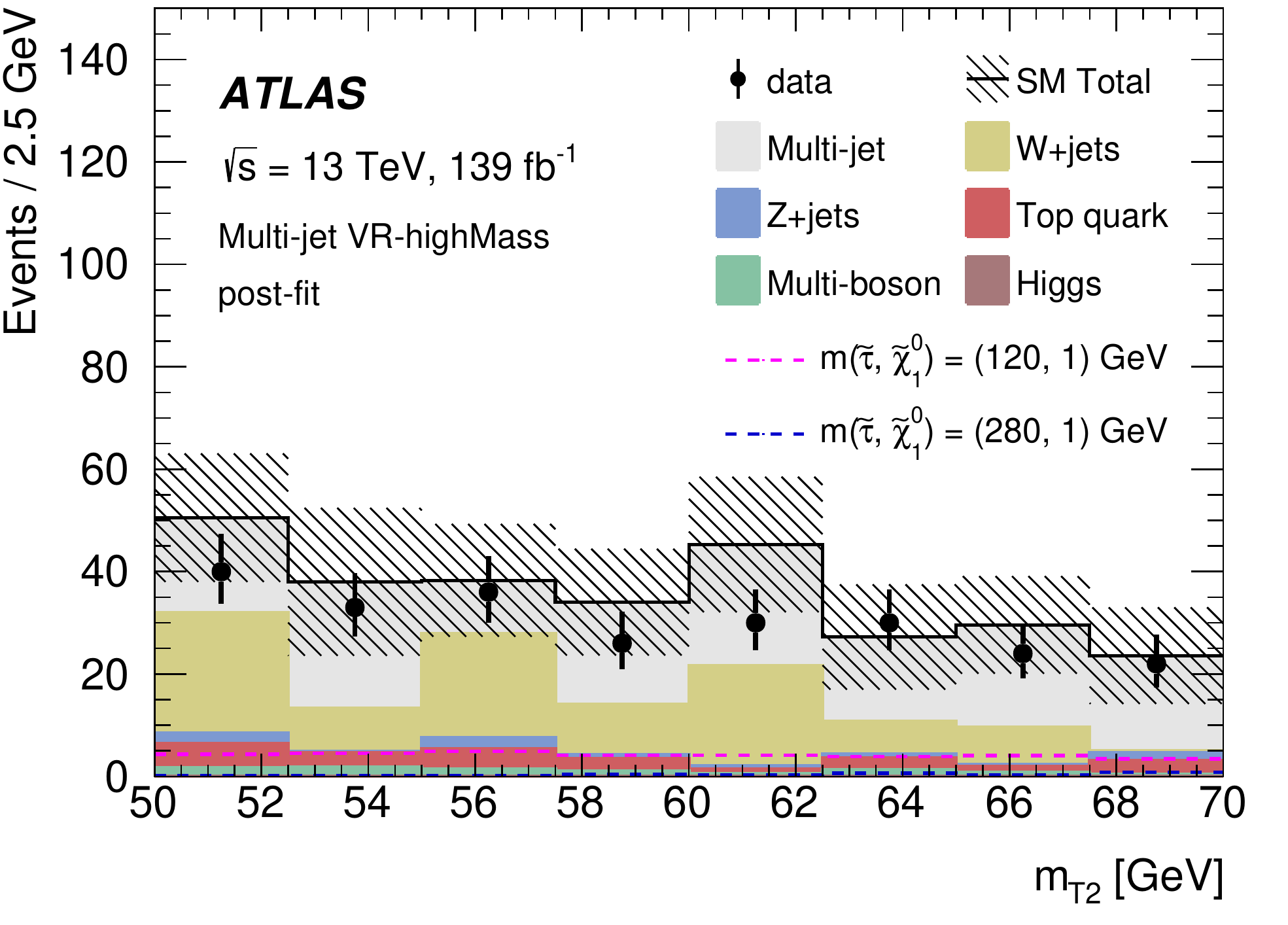}\\
\includegraphics[width=0.495\textwidth]{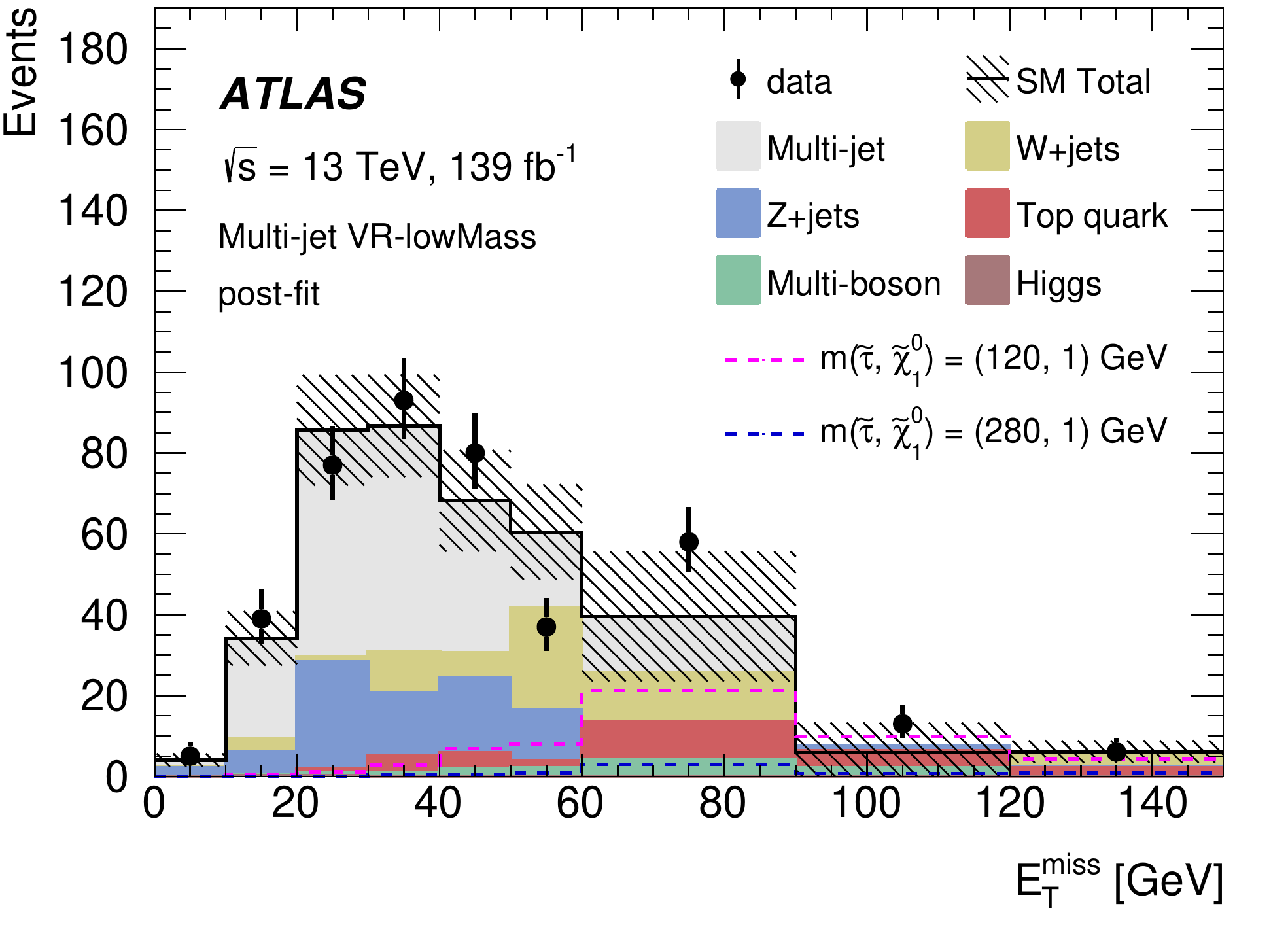}
\includegraphics[width=0.495\textwidth]{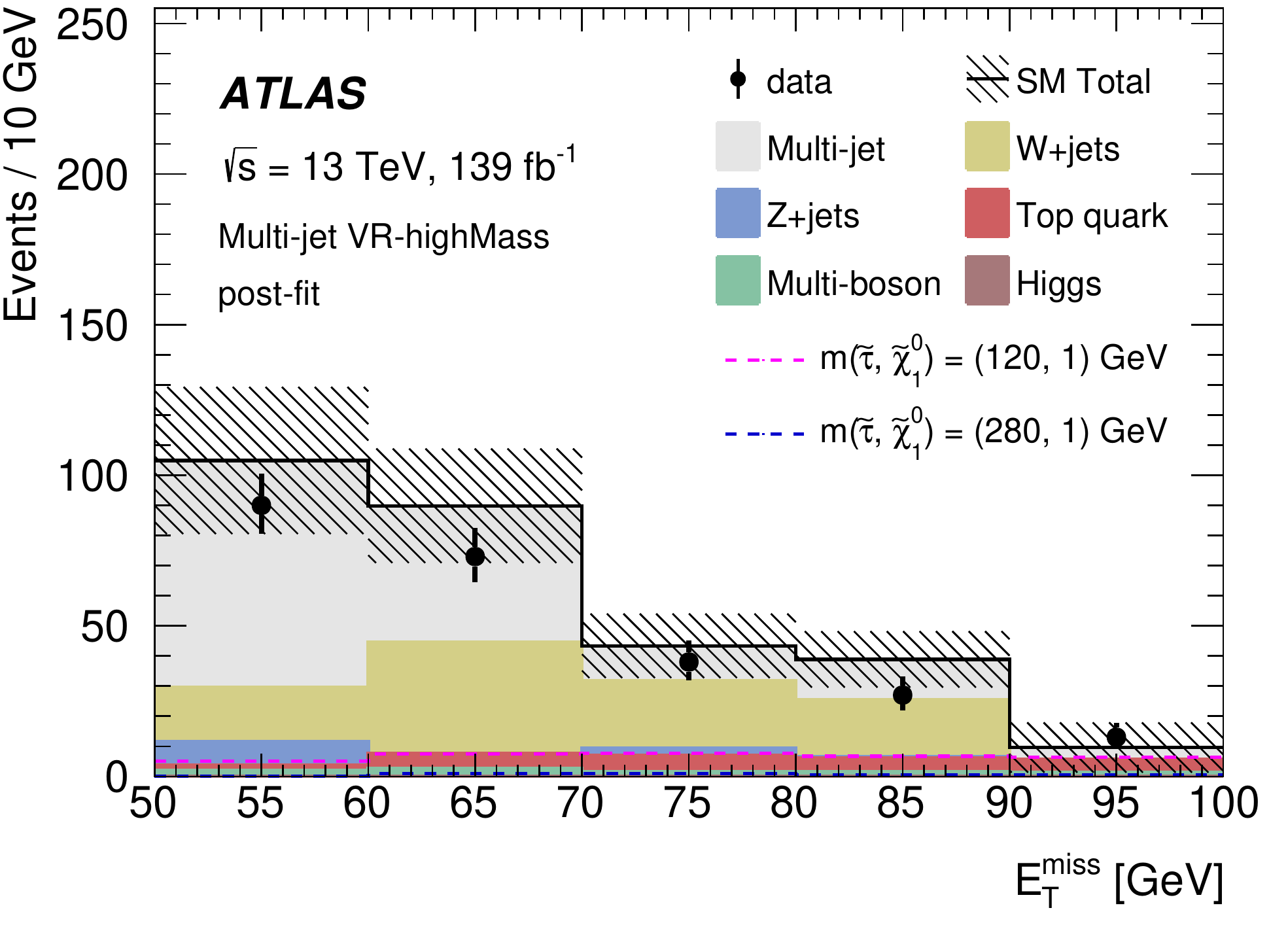}\\
 
\caption{The \mttwo\ (top) and \met (bottom) post-fit distributions in the
multi-jet background validation region VR-F (lowMass) at left and VR-F (highMass) at right.
The stacked histograms show the contribution of each relevant SM process.
The multi-jet shape is taken from VR-E in the ABCD method and the normalization is determined by the transfer factor $T$ and rescaled by a correction factor determined by the background-only fit to data, described in Section \ref{sec:fit}.
The hatched bands represent the combined statistical and systematic uncertainties in the sum of the SM backgrounds shown.
For illustration, the distributions from the SUSY reference points (defined in Section \ref{sec:data})
are also shown as dashed lines.
}
\label{fig:QCD-VR}
\end{figure}

Good agreement between data and the estimated SM background is found for
the \mttwo\ and \met\ distributions in the validation regions, as shown in Figure \ref{fig:QCD-VR}.
 
The contamination from signal in multi-jet CR-A is defined as the
ratio of the number of signal events to the sum of the number of
signal events and SM background processes. It ranges from 0.4\%
(1.2\%) to 9.4\% (21.4\%) for SR-lowMass (SR-highMass) and is taken into account in the fit.
 
The multi-jet background estimation is also validated using a different method,
the \textit{fake-factor method}.
Fake factors (FF) are derived in a region enriched in multi-jet events
consisting of events with two $\tau$-lepton candidates with the same electric
charge. Both $\tau$-lepton candidates are used to compute the fake factor as the
ratio to the number of leading or subleading
$\tau$-lepton candidates satisfying all of the nominal signal identification criteria
over the number of $\tau$-lepton candidates failing the requirement on the jet-veto BDT,
and are parameterized by the $p_\text{T}$, $\eta$ and $N_{\text{prong}}$ of the $\tau$-lepton candidates.
An observed dependence on the BDT output score of the second $\tau$-lepton candidate in the event is also taken into account.
The expected number of multi-jet events entering a selection is computed by applying the fake factors to the yields in a set of sideband regions
where either the leading, subleading or both $\tau$-lepton candidates fail the identification requirements.
The predicted multi-jet event yields
from the ABCD method and the FF method
in both SRs and the multi-jet VRs agree within the statistical and systematic uncertainties.

\subsection{$W$+jets background estimation}
\label{sec:BG_W}
 
The production of $W$+jets events with at least one misidentified $\tau$-lepton
accounts for about 25\% of the
expected SM background in the two SRs.
A dedicated control region ($W$CR) is used to normalize the $W$+jets
MC estimate to data and another region is used to validate
the $W$+jets estimate ($W$ validation region, $W$VR).
To suppress contamination by multi-jet events, the $W$CR is enriched in events
where the $W$ boson decays leptonically into a muon and a
neutrino. Events are selected with a single-muon trigger
using the lowest unprescaled $p_\text{T}$ thresholds available.
Events containing exactly one muon and one candidate $\tau$-lepton with opposite electric charge are selected.
The muon is required to have $p_\text{T}$~$>$ 50 GeV.
The $\tau$-lepton candidate must satisfy the `medium' $\tau$-lepton identification criteria and is required to have $p_\text{T}$~$>$ 60 GeV.

The contribution from events with top quarks is suppressed by
rejecting events containing $b$-tagged jets, and rejecting events
which are kinematically compatible with \ttbar~ production
(top-tagged) through the use of the \textit{contransverse mass} variable $m_{\mathrm {CT}}$ \cite{Tovey:2008}.
The definitions of the $W$CR and $W$VR are given in  Table
\ref{tab:Wregions}. The transverse mass of the $\mu+\met$ system, $ m_{\mathrm{T},\mu}$ is
used to reduce the contribution from $Z$+jets, top-quark, and diboson
events.
To further suppress multi-jet and $Z$+jets events, $\met~$and
$\Delta R(\tau,\mu)$ cuts are applied.
The invariant mass of the muon and $\tau$-lepton, $m(\mu, \tau) $, and
$m_{\mathrm{T},\tau}+m_{\mathrm{T},\mu}$ are used to improve the $W$+jets purity.
Events in the $W$CR ($W$VR) are selected by requiring low (high)
\mttwo.
 
\begin{table}[!htbp]
\caption{Summary of selection requirements for the $W$ control ($W$CR) and validation ($W$VR) regions.
\label{tab:Wregions}}
\small
\centering
\begin{tabular}{c|c}
\toprule
$W$CR    & $W$VR  \\
\hline
\hline
\multicolumn{2}{c}{1 medium $\tau$ and 1 isolated $\mu$ (OS) } \\
\multicolumn{2}{c}{single-muon trigger}\\
\multicolumn{2}{c}{ $p_\text{T}(\tau) >$ 60 GeV, $p_\text{T} (\mu) >$ 50 GeV} \\
\multicolumn{2}{c}{ $\met >$ 60 GeV}   \\
\multicolumn{2}{c}{$b$-jet veto and top-tagged events veto}   \\
\multicolumn{2}{c}{$m(\mu, \tau) >$ 70 GeV } \\
\multicolumn{2}{c}{ 1 $< \Delta R(\mu,\tau) <$ 3.5 } \\
\multicolumn{2}{c}{ 50 $< m_{\mathrm{T},\mu} <$~150~GeV }  \\
\multicolumn{2}{c}{$m_{\mathrm{T},\mu}+m_{\mathrm{T},\tau} >$ 250 GeV }\\
30 $< \mttwo <$ 70 GeV & $\mttwo>$ 70 GeV \\
\hline
\bottomrule
\end{tabular}
\end{table}
 
The multi-jet contribution in the $W$CR ($W$VR) is estimated using the
so-called \textit{OS--SS} method by counting the number of events in data
satisfying the same requirements as for the $W$CR ($W$VR) but with the
electric charge of the two leptons having the same sign.
Events from SM processes other than multi-jet production are
subtracted from the data counts in the SS region using MC simulation.
The OS--SS method relies on the fact that in the multi-jet background
the ratio of SS to OS events is close to unity,
while it is around one-seventh for $W$+jets production.
The latter is dominated by $gu/gd$-initiated processes that often give rise to a jet originating from a quark, the
charge of which is anti-correlated with the $W$ boson charge.
Based on studies with simulated samples, a systematic
uncertainty of 100\% is assigned to the multi-jet estimate in
the $W$CR.
 
\begin{figure}[htbp!]
\centering
\includegraphics[width=0.6\textwidth]{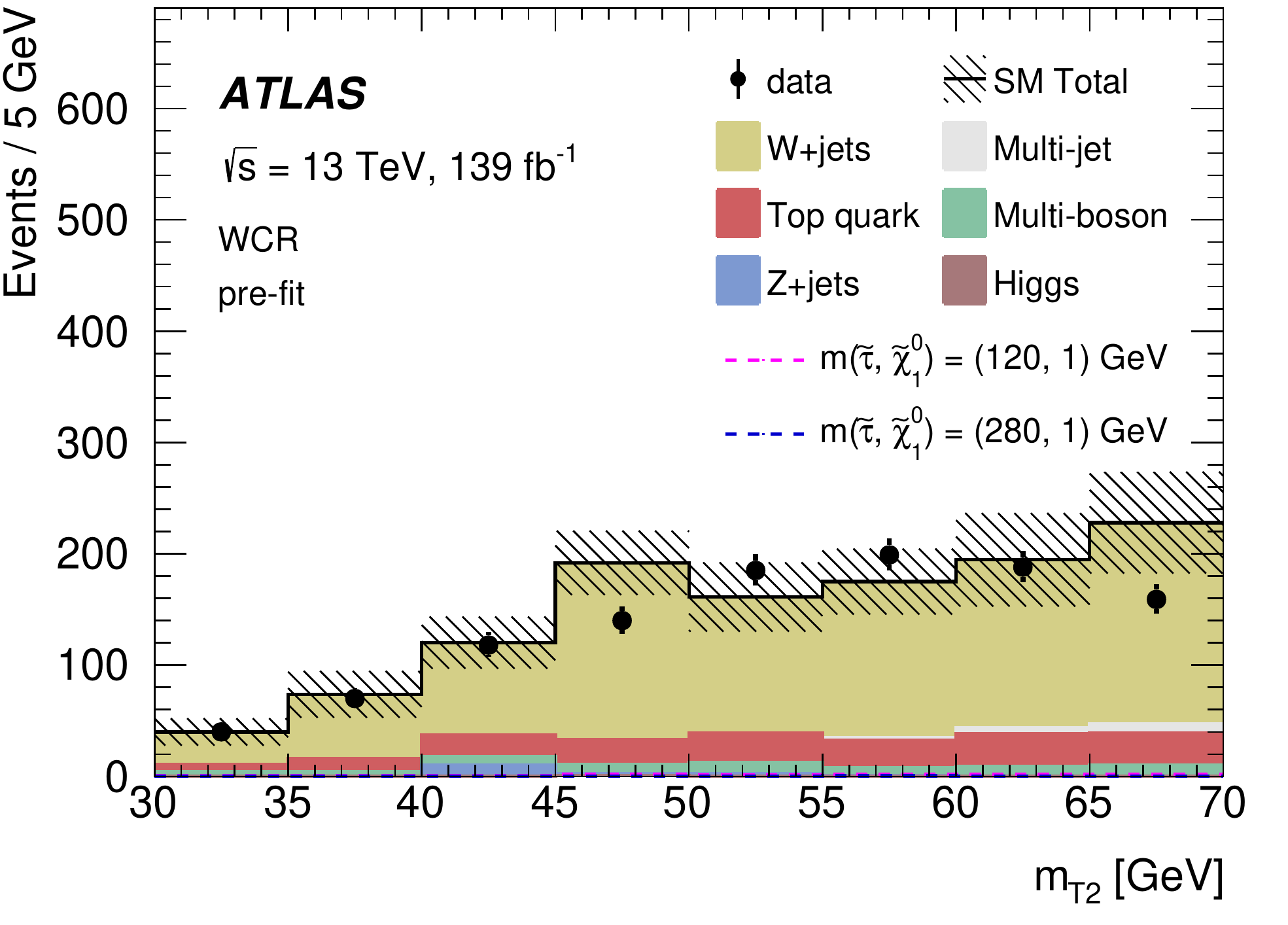}
\caption{
The pre-fit \mttwo~distribution in the $W$CR.
The SM backgrounds other than multi-jet production are estimated from MC simulation.
The multi-jet contribution is estimated from data using the OS--SS method.
The hatched bands represent the combined statistical and systematic uncertainties of the total SM background.
For illustration, the distributions from the SUSY reference points defined in Section \ref{sec:data}
are also shown as dashed lines.
}
\label{fig:WCR}
\end{figure}

The purity of the $W$+jets selection is around 79\% (69\%) in the $W$CR ($W$VR).
The pre-fit \mttwo~distribution in the $W$CR is shown in Figure
\ref{fig:WCR}, and good agreement, both for the normalization and shape, between data and SM predictions is
observed. The contamination from signal in the $W$CR and $W$VR is negligible.
 
\subsection{Irreducible background estimation}
\label{sec:BG_irre}
 
Irreducible SM backgrounds arise mainly from \ensuremath{t\bar{t}},
single top quark, \ensuremath{t\bar{t}}+$V$, $Z$+jets, and multi-boson
(diboson ($WW$, $WZ$ and $ZZ$), triboson ($VVV$)) and Higgs boson.  They
are estimated with MC simulation. Other SM backgrounds are found to be
negligible.
 
The total contribution from \ensuremath{t\bar{t}}, single top quarks, \ensuremath{t\bar{t}}+$V$ and $Z$+jets
amounts to about 8\% and 20\% of the total background in SR-lowMass and SR-highMass, respectively.
The diboson background accounts for 23 -- 25\% of the
total SM contribution in the SRs and mainly arises
from $WW\rightarrow \tau\nu \tau\nu$  and $ZZ \rightarrow \tau\tau
\nu\nu$ events, in which more than 96\% of the contribution is
from events with two real $\tau$-leptons according to the MC simulation.
 
The MC estimates are validated in regions enriched in $Z$+jets,
top-quark, and multi-boson events. For these validation regions, events are required to pass either the combined
di-$\tau$ + \met trigger or the asymmetric di-$\tau$ trigger with the same
offline threshold as described in Table \ref{tab:regions}. Events are required to have
at least two $\tau$-lepton candidates satisfying the `medium' $\tau$-lepton identification
criteria with opposite electric charge, and at least one $\tau$-lepton candidate
must satisfy the `tight' $\tau$-lepton identification criteria in order to be close to the SRs.
In the top-quark validation regions ($T$VR), to increase the
contribution from top-quark events, events must satisfy the
requirement $\Delta R(\tau_{1},\tau_{2}) > 1.2$ and
contain at least one
$b$-tagged jet with $p_{\mathrm T} > 20$ GeV. In order to be close to the
SRs, \mttwo~$>$~60 GeV is required.
In the $Z$+jets and multi-boson validation regions ($Z$VR, $VV$VR), in order to suppress top-quark backgrounds, events containing $b$-tagged jets are vetoed.
To further enhance the purity of $Z$+jets events,
$\Delta R(\tau_{1},\tau_{2})$, $m(\tau_{1},\tau_{2})$ and
\mttwo~requirements are applied.
Further, $m(\tau_{1},\tau_{2})$,
$m_{\mathrm{T},\tau1}+m_{\mathrm{T},\tau2}$ and \mttwo~cuts are
applied in order to enhance multi-boson purity.
The $Z$VRs, $T$VRs and $VV$VRs requirements are summarized in Table \ref{tab:regions}.
 
\begin{figure}[htbp!]
\centering
\includegraphics[width=0.75\textwidth]{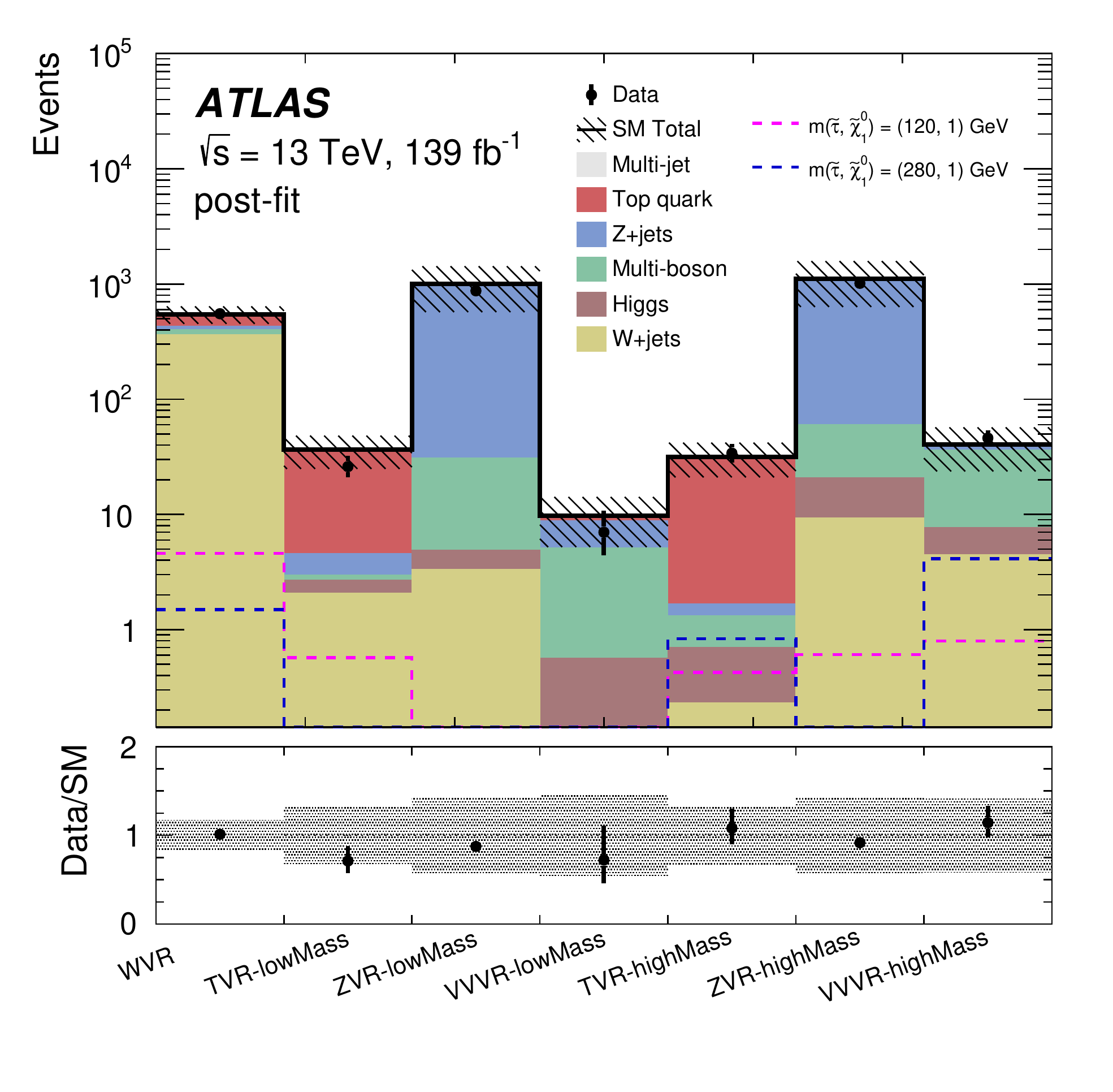}
\caption{The post-fit yields in the $W$VR, $T$VRs, $Z$VRs and $VV$VRs.
The SM backgrounds other than multi-jet production are estimated from MC simulation.
The multi-jet contribution is negligible and is
estimated from data using the ABCD method, using CRs obtained with the
same technique used for the SRs, and described in Section \ref{sec:BG_QCD}. The hatched bands represent
the combined statistical and systematic uncertainties of the total SM background.
The background-only fit to data is used, described in Section \ref{sec:fit}.
For illustration, the distributions from the SUSY reference points (defined in Section \ref{sec:data}) are
also shown as dashed lines. The lower panel shows the ratio of data to
the SM background estimate. 
}
\label{fig:bkg-vr}
\end{figure}

The data event yields and the SM predictions in the $W$+jets ($W$VR), top quark ($T$VR), $Z$+jets ($Z$VR) and
multi-boson ($VV$VR) validation regions are shown in Figure
\ref{fig:bkg-vr}.  The data and SM prediction in each validation region agree within the uncertainties. The purity of the selection in $Z$+jets and \ttbar\ (multi-boson)
events is 83 -- 96\% (47 -- 71\%) in the respective
validation regions.
 
\begin{table}[!htbp]
\caption{Summary of selection requirements for top quark ($T$VR), $Z$+jets ($Z$VR) and
multi-boson ($VV$VR) validation regions.
\label{tab:regions}}
\small
\centering
\begin{tabular}{l||c|c|c|c|c|c}
\toprule
Selections & $T$VR    & $Z$VR &  $VV$VR & $T$VR  &  $Z$VR & $VV$VR \\
& -lowMass    & -lowMass &  -lowMass & -highMass  &  -highMass & -highMass  \\
\hline
\hline
&\multicolumn{6}{c}{$\ge$ 2 medium $\tau$ (OS), $\ge$ 1 tight $\tau$} \\
&$\ge 1~ b$-jet &\multicolumn{2}{c|}{$b$-jet veto} &$\ge 1~ b$-jet &\multicolumn{2}{c}{$b$-jet veto} \\
$m (\tau_{1},\tau_{2}) $&-- & $<$ 70 GeV  &$<$ 110 GeV &-- & $<$ 60 GeV &$<$ 110 GeV\\
$\Delta R(\tau_1,\tau_2) $ &$ >$ 1.2& $<$ 1&-- &$ >$ 1.2& $<$ 1&-- \\
$m_{\mathrm{T},\tau_{1}}+m_{\mathrm{T},\tau_{2}} $&-- &-- & $>$ 250 GeV &-- &-- & $>$ 200 GeV \\
$\mttwo$  &$>$ 60 GeV &$<$ 60 GeV       & $>$ 60 GeV &$>$ 60 GeV &$<$ 60 GeV       & $>$ 60 GeV  \\
\hline
& \multicolumn{3}{c|}{ asymmetric di-$\tau$ trigger} & \multicolumn{3}{c}{ di-$\tau$+\met trigger}   \\
Trigger&\multicolumn{3}{c|}{ 60 $< \met <$ 150 GeV }   & \multicolumn{3}{c}{ $\met >$ 150 GeV}  \\
&\multicolumn{6}{c}{$\tau$ $p_\text{T}$ and \met\ cuts described in Section \ref{sec:SR}}  \\
\hline
\bottomrule
\end{tabular}
\end{table}
 
\section{Systematic uncertainties}
\label{sec:Systematics}
 
Systematic uncertainties have an impact on the estimates of the background and signal event yields
in the control and signal regions.
Uncertainties arising from experimental and theoretical sources are estimated.
 
The main sources of experimental systematic uncertainty in the SM
background estimates include $\tau$-lepton and jet energy scale and resolution,
$\tau$-lepton identification, pileup, and
uncertainties related to the modeling of \met\ in the simulation. The uncertainties in the energy and momentum scale of each of the objects
entering the \met\ calculation are estimated, as well as the uncertainties in the soft-term resolution and scale.
A pileup reweighting procedure is applied to simulation to match the
distribution of the number of reconstructed vertices observed in
data~\cite{Aaboud:2016mmw}. The corresponding uncertainty is derived by a reweighting in which <$\mu$> is varied by $\pm$ 4\%.
The main contributions to experimental systematic uncertainties in the
SRs are from the $\tau$-lepton identification and energy scale~\cite{ATLAS-CONF-2017-029}, 
and jet energy scale and resolution. Systematic uncertainties associated with the $\tau$-lepton triggers are also included with the $\tau$-lepton identification. 
Other contributions are less than 3\%.
 
Theoretical uncertainties affecting the \SHERPA MC event generator
predictions, for the $W$+jets, $Z$+jets and multi-boson samples, are
estimated by varying the PDF sets following the PDF4LHC
recommendations\cite{PDF4LHC} as well as the seven variations of the
QCD renormalization and factorization scales in the matrix element by
factors of 2 and 0.5 avoiding variations in opposite direction\cite{TheorySys}.
For \ttbar\ production, uncertainties in the parton shower simulation,
modeling of initial- and final-state radiation, and associated with
the choice of event generator are considered. The
theory uncertainties are implemented as a separate nuisance parameter
for each scale variation and process; they are not constrained.
The theory uncertainty in $W$+jets (multi-boson) production is mainly
due to the QCD
renormalization scale variation, which amounts to 2 -- 3\% (5 -- 6\%)
compared to the total background yield.
 
The following sources of uncertainty are considered for the ABCD method used to
determine the multi-jet background:  the correlation among the
$\tau$-lepton identification, the charge requirement, and the kinematic variable \mttwo, the
limited number of events in the CRs, and the subtraction of other SM backgrounds.
The systematic uncertainty in the correlation is estimated by comparing the transfer
factor from CR-B to CR-C to that of VR-E to VR-F. The combined experimental
systematic uncertainty and theory uncertainty in the non-multi-jet background subtraction in the control
regions is estimated by considering the systematic
uncertainty of the MC estimates of the non-multi-jet background in these regions.
These uncertainties are negligible due to high multi-jet purity in the CRs.
The statistical uncertainty of the event yields in the control regions is propagated to the signal regions as a systematic uncertainty.

The systematic uncertainties of the background estimates in the SRs are summarized in Table \ref{tab:tauSR_syst}.
The dominant uncertainties in the SRs are the statistical uncertainty of the MC
predictions (11 -- 21\%), $\tau$-lepton identification
and energy scale (10 -- 19\%), and multi-jet background
normalization (8 -- 12\%).
 
The total uncertainty in the signal yields for the SUSY reference
points defined in Section~\ref{sec:data} is about 17 -- 31\%. The dominant uncertainties in the SRs are from $\tau$-lepton identification
and the $\tau$-lepton energy scale (14 -- 29\%) and the statistical uncertainty of the
signal MC
predictions (6 -- 10\%).
The cross-section uncertainty is taken into account as the main source of
the signal modeling theoretical uncertainty;
and it varies from 2\% to 3\% for the SUSY models considered.

\begin{table}[!hbp]
\caption{ The post-fit relative systematic uncertainty (\%) in the background
estimate (signal reference points) in the SR-lowMass and SR-highMass
regions from the leading sources
at top (bottom). Uncertainties from different sources in the background
estimate may be correlated, and do not necessarily add in quadrature to the total
uncertainty.
}
\centering
\smallskip
\smallskip
\begin{tabular}{l|r|r}
\toprule
Source of systematic uncertainty on background prediction& SR-lowMass [\%] & SR-highMass [\%]\\
\hline
\hline
Statistical uncertainty of MC samples & 11 & 21\\
$\tau$-lepton identification and energy scale & 19 & 10 \\
Normalization uncertainties of the multi-jet background & 12 & 8\\
Multi-jet estimation & 4 & 10\\
Jet energy scale and resolution & 5 & 8 \\
Diboson theory uncertainty & 5 & 6 \\
$W$+jets theory uncertainty & 2 & 3 \\
$\met$ soft-term resolution and scale & 2 & 2\\
\hline
Total & 28 & 32 \\
\hline
\hline
 
\hline
Source of systematic uncertainty on signal prediction& SR-lowMass [\%] & SR-highMass [\%]\\
\textit{m} ($\stau$,\ninoone) [GeV]  &  (120, 1) & (280, 1)\\
\hline
\hline
$\tau$-lepton identification and energy scale & 29 & 14 \\
Statistical uncertainty of MC samples & 6 & 10\\
Jet energy scale and resolution & 3 & 2 \\
Signal cross-section uncertainty &2 & 2 \\
$\met$ soft-term resolution and scale & 3 & $<$ 1\\
\hline
Total & 31 & 17 \\
\hline
\bottomrule
 
\end{tabular}\label{tab:tauSR_syst}
\end{table}

\section{Statistical analysis}
\label{sec:fit}
 
The statistical interpretation of the results is performed using the profile likelihood method implemented in the
HistFitter framework~\cite{Baak:2014wma}. The likelihood function is a
product of the probability density functions for every region contributing to the fit.
The number of events in a given CR or SR is described using a Poisson
distribution, the mean of which is the sum of the expected contributions from all background and signal sources.
The systematic uncertainties in the expected event yields are included
as nuisance parameters and are assumed to follow a Gaussian distribution with a width determined by the size of the uncertainty.
Correlations between control and signal regions, and among background processes are taken into account through common nuisance
parameters. The fit parameters are determined by maximizing the
product of the Poisson probability functions and the Gaussian
probability constraints for the nuisance parameters.

Three types of fits are
performed for the combined SR-lowMass and SR-highMass regions.
 
\begin{itemize}
\item
The \textit{background-only} fit uses as input the number of observed events and expected SM contributions in the multi-jet CR-A and $W$CR
as well as the transfer factors, which relate the number of multi-jet
or $W$+jets events in the control regions to the number in the signal regions.
The free parameters in the fit are the normalizations of the $W$+jets and multi-jet contributions.
The signal is assumed to be absent in this fit.
\item
A \textit{model-independent limit} fit combines the data event yield in a given SR with the SM background estimate
and its uncertainties to test whether any non-SM signal contributes to the SR.
The significance of a possible excess of observed events over the SM prediction
is quantified by the one-sided probability, $p (\text{signal} = 0)$ denoted by $p_0$, of the background alone to fluctuate to the observed number of events or higher
using the asymptotic formula described in Ref.~\cite{statforumlimitsl}. The background yields and uncertainties are taken from the background-only fit results.
The presence of a non-SM signal would manifest itself as a small $p_0$ value.
\item In the \textit{model-dependent limit} fit the SUSY signal is allowed to populate both the signal regions and the control regions, and it is
scaled by a freely floating signal normalization factor.
The background normalization factors are also determined simultaneously in the fit.
A SUSY model with a specific set of sparticle masses is rejected if
95\% CL upper limit on the signal normalization factor obtained in this fit is smaller than unity.
\end{itemize}

\section{Results}
\label{sec:result}
 
\begin{figure}[htbp!]
\centering
\includegraphics[width=0.49\textwidth]{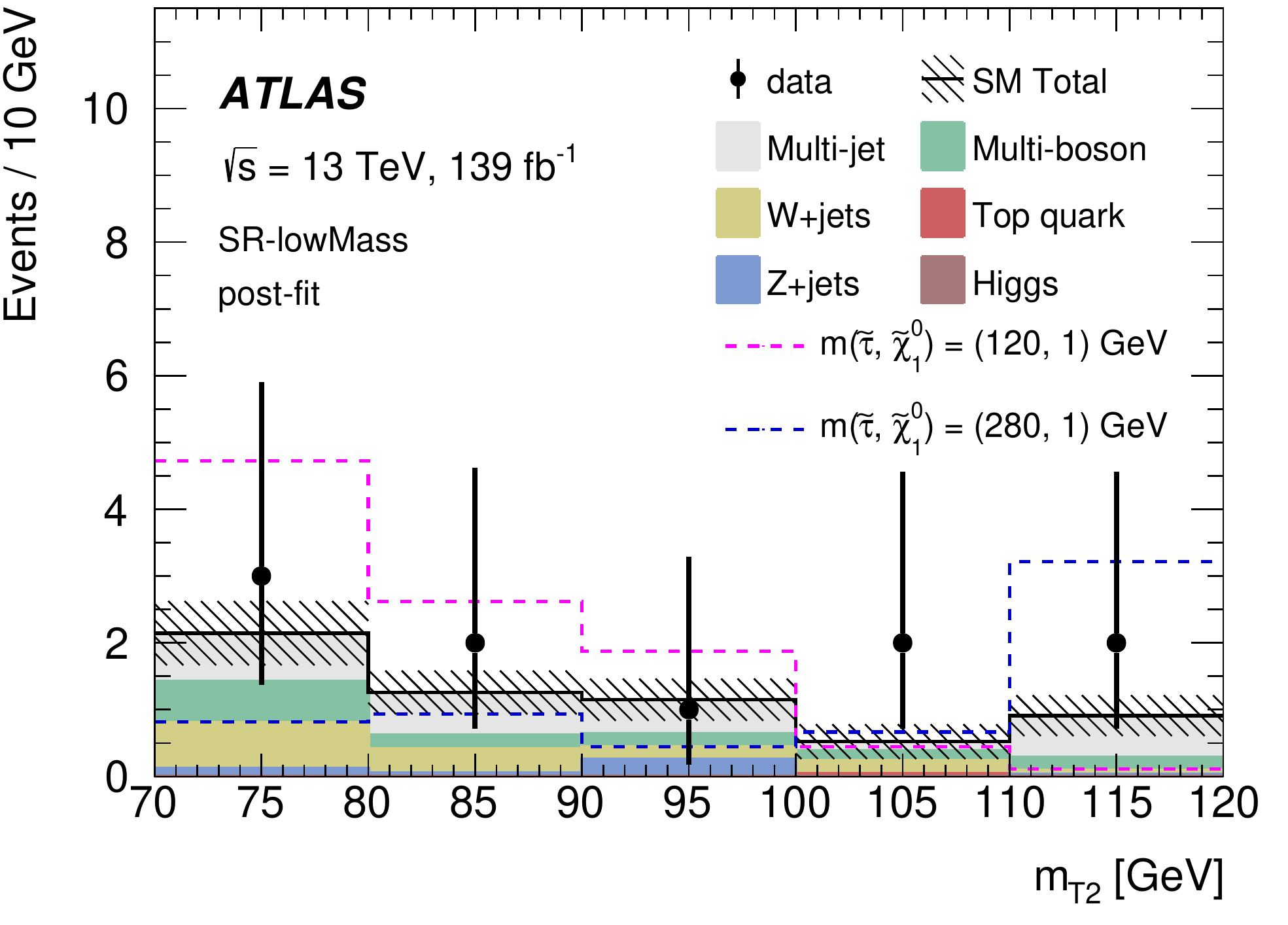}
\includegraphics[width=0.49\textwidth]{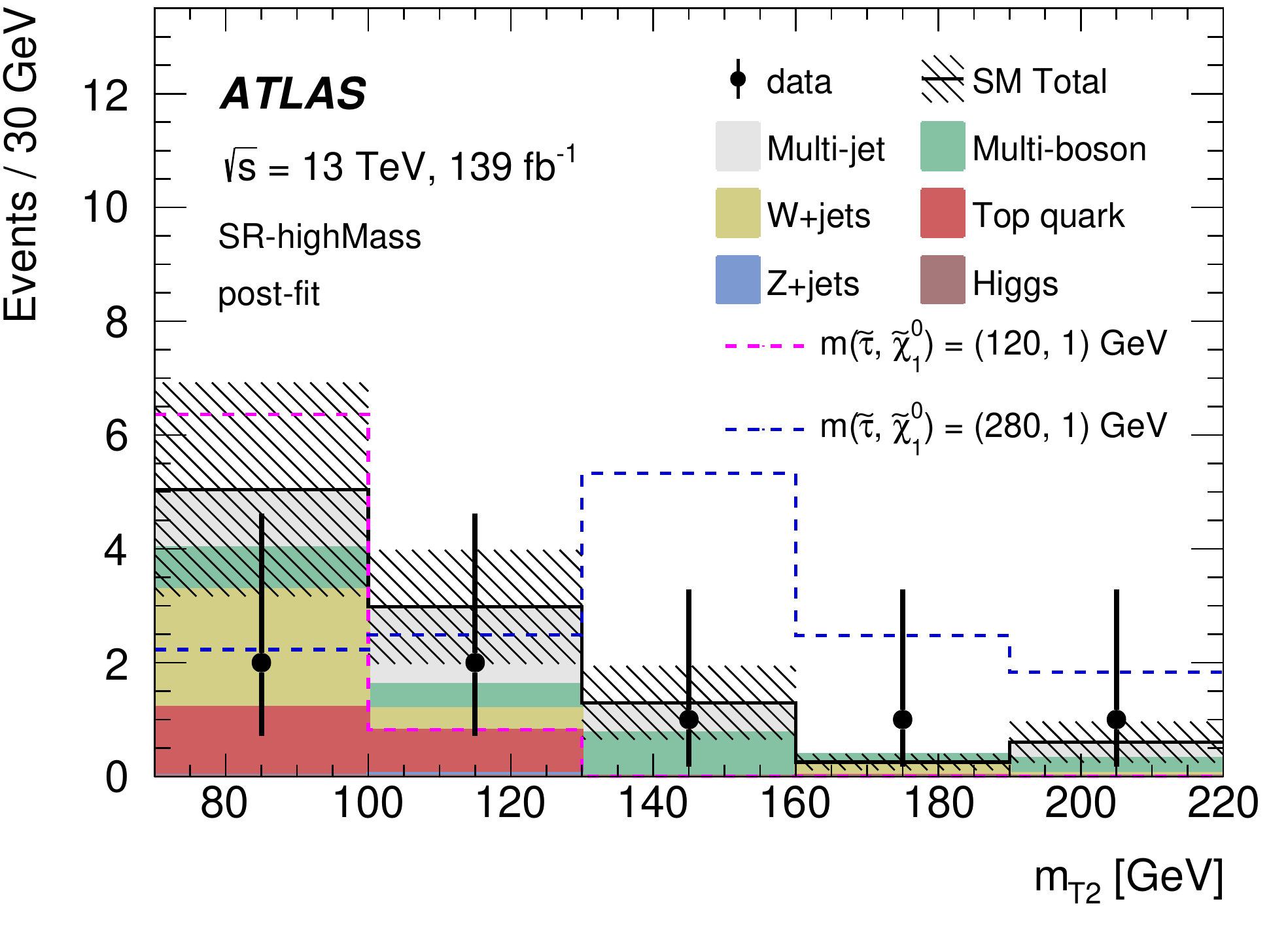}\\
\caption{
The post-fit \mttwo\ distribution for SR-lowMass (left) and SR-highMass (right).
The stacked histograms show the expected SM backgrounds.
The multi-jet contribution is estimated from data using the ABCD method.
The contributions of multi-jet and $W$+jets events are scaled with the
corresponding normalization factors derived from the background-only fit described in Section \ref{sec:fit}. The hatched bands represent the sum in quadrature of systematic and statistical uncertainties of the total SM background.
For illustration, the distributions from the SUSY reference points (defined in Section \ref{sec:data}) are
also shown as dashed lines.  The last bin includes the overflow events.
}
\label{fig:SR-result}
\end{figure}

The observed number of events in each control, and signal region and the expected contributions from SM processes
are given in Table~\ref{tab:results}. The contributions of multi-jet and $W$+jets events are scaled with the normalization factors
obtained from the background-only fit described in Section~\ref{sec:fit}.
The multi-jet normalization in the SR is
1.03 times the prediction from the ABCD method and has an uncertainty of around 28\%, due to
the small number of observed events in the multi-jet CR-A.
The $W$+jets normalization is 0.91$\pm$0.12.
The \mttwo\ distributions
are shown in Figure \ref{fig:SR-result} for data, expected SM backgrounds, and
the SUSY reference points defined in Section~\ref{sec:data}.
In both signal regions, observations and background predictions are
found to be compatible within uncertainties.
 
Individual model-independent fits of the SR-lowMass and SR-highMass are used to derive
the one-sided $p_0$-values.
In addition, the observed and expected 95\% CL upper limits on the visible non-SM cross section ($\sigma_{\mathrm{vis}}^{95}$)
are derived using the same model-independent fits.
The $\sigma_{\mathrm{vis}}^{95}$ is defined as the product of acceptance,
reconstruction efficiency and production cross section, obtained normalizing the limit on the non-SM
yield in the SR by the integrated luminosity of the data sample.
These results are shown in Table~\ref{tab:results_sig}.
All limits are calculated using the \CLs prescription~\cite{CLs:2002}.
 
\begin{table}[htbp!]
\centering
\caption{
Observed and expected numbers of events in the control and signal regions.
The expected event yields of SM processes are given after the background-only fit described in Section~\ref{sec:fit}.
The entries marked as ``--" are negligible.
The uncertainties correspond to the sum in quadrature of statistical  and systematic uncertainties.
The correlation of systematic uncertainties among control regions and
among background processes is fully taken into account.
}
\smallskip
\smallskip
\renewcommand{\arraystretch}{1.3}
\begin{tabular}{l|r|r|r|r|r}
\toprule
SM process   & Multi-jet CR-A& Multi-jet CR-A & $W$CR  & SR & SR                      \\
& -lowMass    & -highMass &   &  -lowMass & -highMass  \\
\hline
\hline
 
Diboson      & $1.4 \pm 0.6$   & $1.9\pm 1.0$  & $63 \pm 21$    & $1.4 \pm 0.8$  & $2.6 \pm 1.4$  \\
$W$+jets     & $13 \pm 4$   & $4_{-4}^{+7}$    & $850 \pm 70$     & $1.5 \pm 0.7$  & $2.5 \pm1.8$    \\
Top quark      & $2.7 \pm 0.9$  & $3.3 \pm 1.6$ & $170 \pm 40$     & $0.04_{-0.04}^{+0.80}$    & $2.0 \pm 0.6$   \\
$Z$+jets     & $0.25_{-0.25}^{+1.43}$ & $1.5 \pm 0.8$ & $13 \pm 7$  & $0.4_{-0.4}^{+0.5}$ & $0.05_{-0.05}^{+0.13}$  \\
Multi-jet     & $55 \pm 10$  & $16 \pm 6$  & --  & $2.6 \pm 0.7$  & $3.1 \pm 1.4$  \\
\hline
SM total         & $72 \pm 8$   & $27 \pm 5$  & $1099 \pm 33$     & $6.0 \pm 1.7$    & $10.2 \pm 3.3$    \\
\hline
Observed         & $72$              & $27$              & $1099$               & $10$              & $7$   \\
 
\hline
\bottomrule
\end{tabular}
\label{tab:results}
\end{table}

\begin{table}[htbp!]
\centering
\caption{
Expected event yields for the SUSY reference points (defined in
Section~\ref{sec:data}) are shown along with the one-sided
$p_0$-values, and the observed and expected 95\% CL
upper limits on the visible non-SM cross section ($\sigma_{\mathrm
{vis}}^{95}$) are given. Values of $p_0>0.5$  are truncated at $p_0=0.5$.
}
\smallskip
\smallskip
\renewcommand{\arraystretch}{1.3}
\begin{tabular}{l|r|r}
\toprule
&  SR-lowMass & SR-highMass    \\
\hline
$m$ ($\stau$,\ninoone)  = (120, 1) GeV &   $9.8 \pm 3.0$  &   $7.2 \pm 2.2$  \\
$m$ ($\stau$,\ninoone)  = (280, 1) GeV &   $6.1 \pm 1.5$  &   $14.4 \pm 2.5$  \\
\hline
\rule{0pt}{2.6ex}
$p_0$                                                    &0.11       &0.50      \\
Expected $\sigma_\text{vis}^{95}$ [fb]  & $ { 0.055 }^{ +0.025 }_{ -0.014 }$ & $ { 0.065 }^{ +0.025 }_{ -0.019 }$ \\ [+1mm]
Observed $\sigma_\text{vis}^{95}$ [fb]  & $0.08$ & $0.05$   \\
\hline
\hline
\bottomrule
\end{tabular}
\label{tab:results_sig}
\end{table}

\section{Interpretation}
\label{section:interpretation}
 
\begin{figure}[htbp!]
\centering
\includegraphics[width=0.49\textwidth]{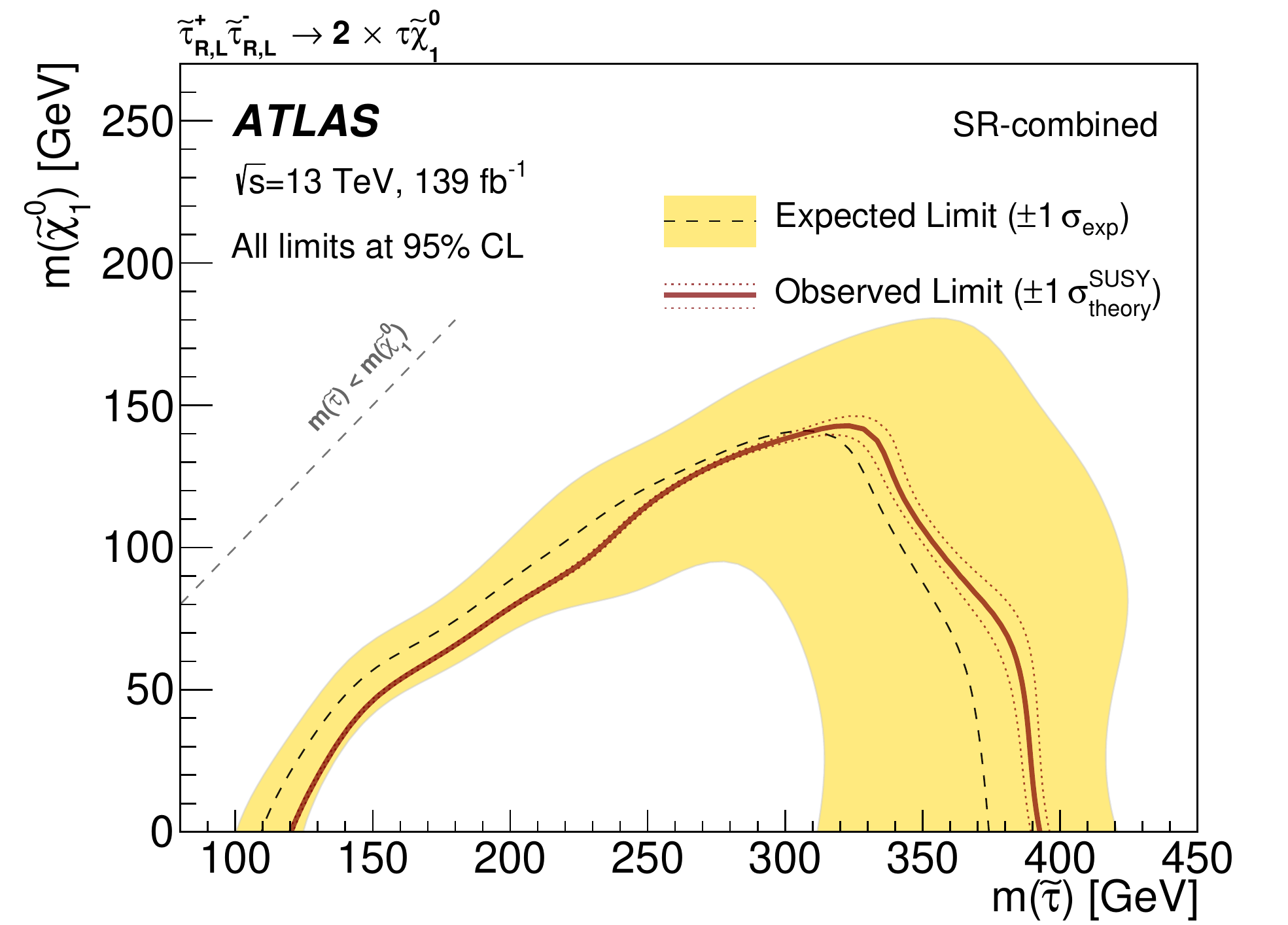}
\includegraphics[width=0.49\textwidth]{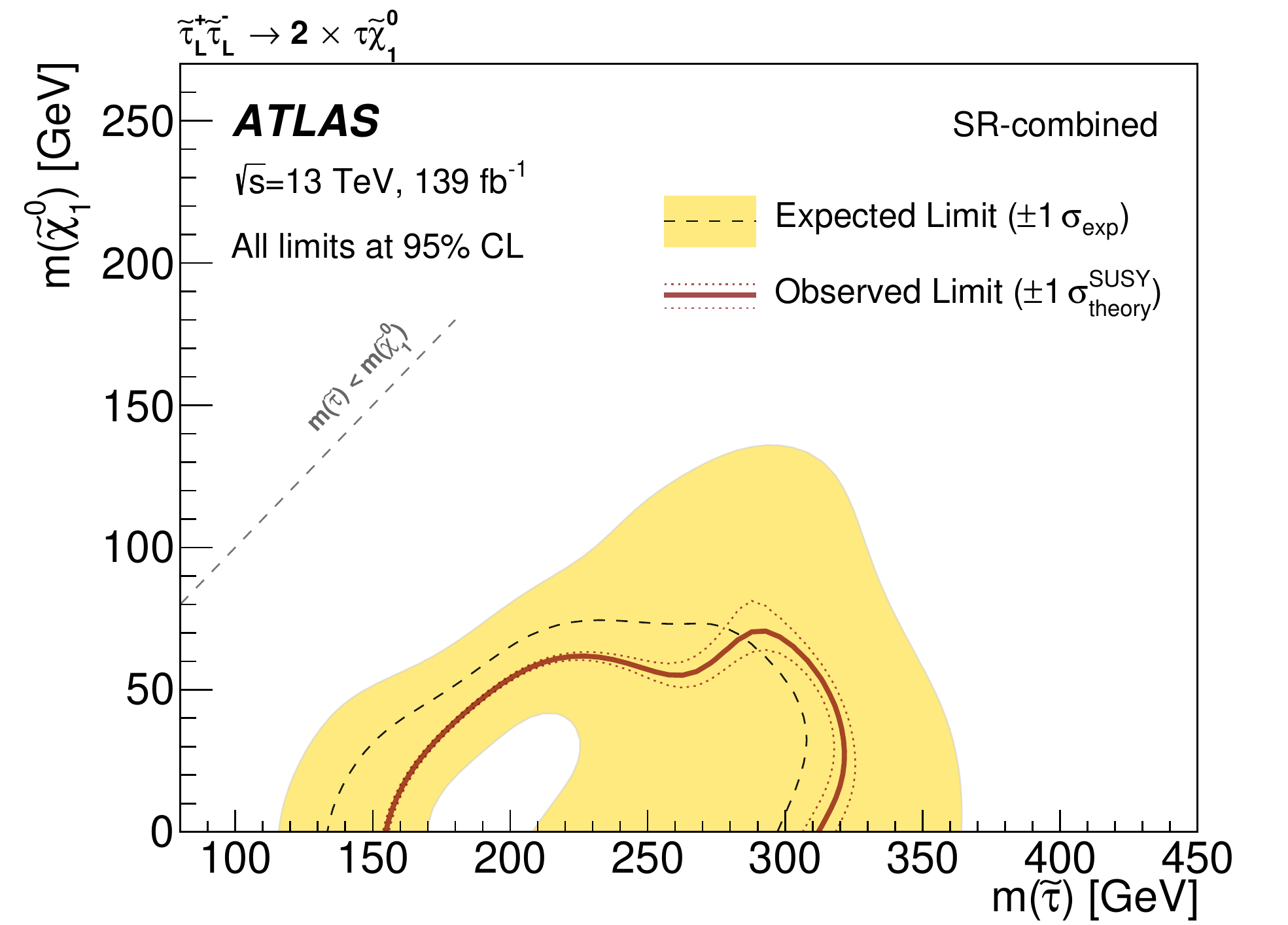}
\caption{The 95\% CL exclusion contours for the combined
fit of SR-lowMass and SR-highMass for simplified
models with  (a) combined $\stau^{+}_\mathrm{R,L} \stau^{-}_\mathrm{R,L}$ production and (b) $\stau^{+}_\mathrm{L} \stau^{-}_\mathrm{L}$ only production.
The text provides details of exclusion curves and uncertainty
bands.
}
\label{fig:alimit}
\end{figure}

In the absence of a significant excess over the expected SM background, the observed and expected numbers of events in the signal regions are used to
set exclusion limits at 95\% CL using the
model-dependent limit fit.
The exclusion limits for the combined SR-lowMass and SR-highMass for the simplified models described in Section \ref{sec:data}
are shown in Figure \ref{fig:alimit}.
The solid (dashed) lines show the observed (expected)
exclusion contours. The band around the expected limit shows the
$\pm1\sigma$ variations, including all
uncertainties except theoretical uncertainties in the signal cross section.
The dotted lines around the observed limit indicate the sensitivity to
$\pm1\sigma$ variations of the theoretical uncertainties in the signal cross section.
 
Stau masses from 120 GeV to 390 GeV are excluded for a massless lightest
neutralino in the scenario of combined stau-left and stau-right  $\stau^{+}_\mathrm{R,L} \stau^{-}_\mathrm{R,L}$ production.
For stau-left pair production only,  $\stau^{+}_\mathrm{L} \stau^{-}_\mathrm{L}$, the
exclusion region extends from 155 GeV to 310 GeV. While the stau-left
pairs have a higher production cross section, the stau-right
pairs have a higher efficiency times acceptance due to kinematic differences in the resulting decay products.

These limits extend significantly beyond previous
results~\cite{SUSY-2014-05, CMS-SUS-18-006, CMS-SUS-13-006, CMS-SUS-17-003} in the high $\stau$ mass
region.
 
Uncertainties in each background from scale variations are fully
correlated across regions and bins, and uncorrelated between
processes. In some cases this may result in uncertainties canceling
out, while the higher-order corrections may not cancel out.
A different fit with scale variations uncorrelated in all bins results
in a less than 1\% change in the limits on the excluded cross section
near the edge of exclusion for combined left and right $\stau$ $\stau$ production.

\FloatBarrier
\section{Conclusion}
 
Searches for stau-pair production in events with
at least two hadronically decaying $\tau$-leptons and missing
transverse momentum were performed using \ourintlumi of
$pp$ collision data at $\sqrt{s}=13$~TeV recorded with the ATLAS
detector at the LHC.
Agreement between data and SM predictions is observed in two optimized
signal regions. The results are used to set limits on the visible cross section
for events beyond the Standard Model in each signal region.
 
Exclusion limits are placed on parameters of simplified electroweak
supersymmetry models in scenarios of stau-pair production. 
Stau masses from 120 GeV to 390 GeV are excluded for a massless lightest
neutralino in the scenario of direct production of
stau pairs, with each stau decaying into the lightest neutralino and
one $\tau$-lepton. These limits extend significantly beyond previous
results by the ATLAS and CMS experiments in the high $\stau$ mass
region.
 
\section*{Acknowledgments}

We thank CERN for the very successful operation of the LHC, as well as the
support staff from our institutions without whom ATLAS could not be
operated efficiently.
 
We acknowledge the support of ANPCyT, Argentina; YerPhI, Armenia; ARC, Australia; BMWFW and FWF, Austria; ANAS, Azerbaijan; SSTC, Belarus; CNPq and FAPESP, Brazil; NSERC, NRC and CFI, Canada; CERN; CONICYT, Chile; CAS, MOST and NSFC, China; COLCIENCIAS, Colombia; MSMT CR, MPO CR and VSC CR, Czech Republic; DNRF and DNSRC, Denmark; IN2P3-CNRS and CEA-DRF/IRFU, France; SRNSFG, Georgia; BMBF, HGF and MPG, Germany; GSRT, Greece; RGC and Hong Kong SAR, China; ISF and Benoziyo Center, Israel; INFN, Italy; MEXT and JSPS, Japan; CNRST, Morocco; NWO, Netherlands; RCN, Norway; MNiSW and NCN, Poland; FCT, Portugal; MNE/IFA, Romania; MES of Russia and NRC KI, Russia Federation; JINR; MESTD, Serbia; MSSR, Slovakia; ARRS and MIZ\v{S}, Slovenia; DST/NRF, South Africa; MINECO, Spain; SRC and Wallenberg Foundation, Sweden; SERI, SNSF and Cantons of Bern and Geneva, Switzerland; MOST, Taiwan; TAEK, Turkey; STFC, United Kingdom; DOE and NSF, United States of America. In addition, individual groups and members have received support from BCKDF, CANARIE, Compute Canada and CRC, Canada; ERC, ERDF, Horizon 2020, Marie Sk{\l}odowska-Curie Actions and COST, European Union; Investissements d'Avenir Labex, Investissements d'Avenir Idex and ANR, France; DFG and AvH Foundation, Germany; Herakleitos, Thales and Aristeia programmes co-financed by EU-ESF and the Greek NSRF, Greece; BSF-NSF and GIF, Israel; CERCA Programme Generalitat de Catalunya and PROMETEO Programme Generalitat Valenciana, Spain; G\"{o}ran Gustafssons Stiftelse, Sweden; The Royal Society and Leverhulme Trust, United Kingdom.
 
The crucial computing support from all WLCG partners is acknowledged gratefully, in particular from CERN, the ATLAS Tier-1 facilities at TRIUMF (Canada), NDGF (Denmark, Norway, Sweden), CC-IN2P3 (France), KIT/GridKA (Germany), INFN-CNAF (Italy), NL-T1 (Netherlands), PIC (Spain), ASGC (Taiwan), RAL (UK) and BNL (USA), the Tier-2 facilities worldwide and large non-WLCG resource providers. Major contributors of computing resources are listed in Ref.~\cite{ATL-GEN-PUB-2016-002}.
Unfortunately, our colleague Jihyun Jeong passed away shortly before the completion of this work.
 
 
\FloatBarrier
 
\printbibliography

\newpage
 
\clearpage \input{atlas_authlist}
\end{document}

%% file: atlas_authlist.tex
 
\begin{flushleft}
{\Large The ATLAS Collaboration}

\bigskip

G.~Aad$^\textrm{\scriptsize 102}$,    
B.~Abbott$^\textrm{\scriptsize 129}$,    
D.C.~Abbott$^\textrm{\scriptsize 103}$,    
A.~Abed~Abud$^\textrm{\scriptsize 71a,71b}$,    
K.~Abeling$^\textrm{\scriptsize 53}$,    
D.K.~Abhayasinghe$^\textrm{\scriptsize 94}$,    
S.H.~Abidi$^\textrm{\scriptsize 167}$,    
O.S.~AbouZeid$^\textrm{\scriptsize 40}$,    
N.L.~Abraham$^\textrm{\scriptsize 156}$,    
H.~Abramowicz$^\textrm{\scriptsize 161}$
% ...
\bigskip
\\

% ...
$^{40}$Niels Bohr Institute, University of Copenhagen, Copenhagen; Denmark.\\
% ...
$^{53}$II. Physikalisches Institut, Georg-August-Universit\"{a}t G\"ottingen, G\"ottingen; Germany.\\
% ...
$^{71}$$^{(a)}$INFN Sezione di Pavia;$^{(b)}$Dipartimento di Fisica, Universit\`a di Pavia, Pavia; Italy.\\
% ...
$^{94}$Department of Physics, Royal Holloway University of London, Egham; United Kingdom.\\
% ...
$^{102}$CPPM, Aix-Marseille Universit\'e, CNRS/IN2P3, Marseille; France.\\
$^{103}$Department of Physics, University of Massachusetts, Amherst MA; United States of America.\\
% ...
$^{129}$Homer L. Dodge Department of Physics and Astronomy, University of Oklahoma, Norman OK; United States of America.\\
% ...
$^{156}$Department of Physics and Astronomy, University of Sussex, Brighton; United Kingdom.\\
% ...
$^{161}$Raymond and Beverly Sackler School of Physics and Astronomy, Tel Aviv University, Tel Aviv; Israel.\\
% ...
$^{167}$Department of Physics, University of Toronto, Toronto ON; Canada.\\
% ...
$^{a}$ Also at Borough of Manhattan Community College, City University of New York, New York NY; United States of America.\\
$^{b}$ Also at CERN, Geneva; Switzerland.\\
% ...

\end{flushleft}
